\documentclass[english]{article}%
\usepackage{amssymb}
\usepackage{fontenc}
\usepackage[latin1]{inputenc}
\usepackage{amsmath}
\usepackage{fontenc}
\usepackage{babel}
\usepackage{amsfonts}
\usepackage{graphicx}
\usepackage{geometry}
\usepackage{cite}%
\providecommand{\U}[1]{\protect\rule{.1in}{.1in}}
 
\begin{document}

\title{{\Large Pair production from the vacuum by a weakly inhomogeneous 
space-dependent electric potential step}}
\author{S. P.\ Gavrilov$^{b,d}$\thanks{e-mail: gavrilovsergeyp@yahoo.com,
gavrilovsp@herzen.spb.ru}, D.\ M.\ Gitman$^{a,c,d}$\thanks{e-mail:
gitman@if.usp.br}, and A.\ A.\ Shishmarev$^{c}$\thanks{e-mail:
a.a.shishmarev@mail.ru}\\$^{a}$ \emph{P.N. Lebedev Physical Institute, 53 Leninskiy prospect, 119991
Moscow, Russia;}\\$^{b}$ \emph{Department of General and Experimental Physics, }\\\emph{Herzen State Pedagogical University of Russia,}\\\emph{Moyka embankment 48, 191186 St. Petersburg, Russia;}\\$^{c}$ \emph{Institute of Physics, University of S{\~{a}}o Paulo, }\\\emph{Rua do Mat{\~{a}}o, 1371, CEP 05508-090, S\~{a}o Paulo, SP, Brazil }\\$^{d}$ \emph{Department of Physics, Tomsk State University, 634050 Tomsk,
Russia;}}
\maketitle

\begin{abstract}
There exists a clear physical motivation for theoretical studies of the vacuum
instability related to the production of electron-positron pairs from a vacuum
due to strong external electric fields. Various nonperturbative (with respect
to the external fields) calculation methods were developed. Some of these
methods are based on possible exact solutions of the Dirac equation.
Unfortunately, there are only few cases when such solutions are known.
Recently, an approximate but still nonperturbative approach to treat the
vacuum instability caused by slowly varying $t$-electric potential steps (time
dependent external fields that vanish as $|t|\rightarrow\infty$), which does
not depend on the existence of the corresponding exact solutions, was
formulated in Ref. [S. P. Gavrilov, D. M. Gitman, Phys. Rev. D \textbf{95},
076013 (2017)]. Here, we present an approximate
calculation method to treat nonperturbatively the vacuum instability in
arbitrary weakly inhomogeneous $x$-electric potential steps (time-independent
electric fields of a constant direction that are concentrated in
restricted space areas, which means that the fields vanish as $|x|\rightarrow
\infty$) in the absence of the corresponding exact solutions. Defining the
weakly inhomogeneous regime in general terms, we demonstrate the universal
character of the vacuum instability. This universality is associated with a
large density of states excited from the vacuum by the electric field. Such a
density appears in our approach as a large parameter. We derive universal
representations for the total number and current density of the created particles.
Relations of these representations with a locally constant field approximation
for Schwinger's effective action are found.

\emph{Keywords}: Particle creation, Dirac equation, constant inhomogeneous
external field, Schwinger effect \\
PACS number(s): 12.20.Ds, 12.20.-m

\end{abstract}

\section{Introduction}

Since the work of Schwinger pointing to the possible vacuum instability due to
the pair production in strong external electriclike fields (the Schwinger
effect) \cite{Schwinger51}, this effect has always attracted the attention of
physicists. At present we know that astrophysical objects such as black holes
and hot strange stars can generate huge electromagnetic fields in their
vicinity (dozens of times higher than Schwinger's critical field, $E_{c}%
=m^{2}/e$), see, e.g. Refs. \cite{Usov97,Alf+etal01,Ruf+etal11,Web+etal14}. It
can also be seen that in some situations in graphene and similar
nanostructures the vacuum instability effects caused by strong (with respect to massless fermions) electric fields are of significant interest; see, e.g.,
Refs.
\cite{GelTan16,allor08,GavGitY12,VafVish14,KaneLM15,Olad+etal17,Akal+etal19}
and references therein. Recent progress in laser physics allows one to hope that the particle creation effect will be experimentally observed in laboratory conditions in the near future, as the strong laser experimental community, for example, Center for Relativistic Laser Science (CoReLS), Extreme Light Infrastructure (ELI), and Exawatt Center for Extreme Light Studies (XCELS), is slowly approaching the critical field strengths for observable pair production (see Ref. \cite{LaserRev} for a review).
Thus, there exists a clear physical motivation for
theoretical studies of the vacuum instability. Firstly, it seems necessary to
us to mention theoretical works devoted to various nonperturbative (with
respect to the external field) calculation methods. Some of these methods are
formulated for time-dependent external fields that vanish as
$|t|\rightarrow\infty$ (for $t$-electric potential steps in what follows) and
are based on possible exact solutions of the Dirac equation; see, e.g.,
\cite{FGS,Nikis79,General1,Ruf10,GelTan16}. Some of the methods are based on the
analysis of the Schwinger effective action (see \cite{Dunn04} for a
review). The so-called derivative expansion approximation method,
being applied to the Schwinger effective action, allows one to treat
effectively arbitrary slowly varying in time strong fields
\cite{DunnH98,GusSh99}. We note that the locally constant field approximation
(LCFA), which is to limit oneself to leading contributions of the derivative
expansion of the effective action, allows for reliable results for
electromagnetic fields of arbitrary strength; see, for example, Refs.~\cite{GalN83,GiesK17}. An alternative approach to treat slowly varying
$t$-electric potential steps, which does not depend on the existence of the
corresponding exact solutions, was formulated in Ref. \cite{GavGit17}.

When achieving extreme field strengths, the inhomogeneity of realistic
external fields becomes important. At the same time in astrophysics and in the
physics of graphene, electric fields can be considered as time independent;
see, e.g., references cited above. Thus, it is important to study the
effect of pair creation in strong constant inhomogeneous fields and to
develop corresponding nonperturbative methods. Here, it is natural to start
with considering the vacuum instability caused by time-independent
inhomogeneous electric fields of a constant direction that depend on only one
coordinate $x$ and are concentrated in restricted space areas, which means
that the fields vanish as $|x|\rightarrow\infty$. The latter fields represent
a kind of so-called $x$-electric potential steps for charged particles.
Nonperturbative methods for treating quantum effects in $t$-electric potential
steps with the help of exact solutions of the Dirac equation are not directly
applicable to the $x$-electric potential steps. One of the main differences
between these approaches is that unlike the case of $t$-electric potential steps, the
magnitude of the corresponding $x$-potential step, $\Delta U$, is crucial for
pair creation from a vacuum regardless of a field intensity. Only critical steps,
$\Delta U>2m$ ($m$ is electron mass), produce electron-positron pairs and this
production occurs only in a finite range of quantum numbers that is called the
Klein zone. Depending on the localization of the constant field, a critical
point (critical surface) exists in the space of inhomogeneous electric field
configurations where the pair production probability vanishes. This
critical surface separates the Klein zone from the adjacent ranges where the pair
production is impossible. Note that the absence of the critical surface in the
case of the $t$-electric potential step arises as a consequence of neglecting
the fact that a realistic electric field occupies a finite space region.

An adequate nonperturbative technique for treating the vacuum instability in
the $x$-electric potential steps was elaborated on in Ref. \cite{x-case}. Similar to
the case of $t$-electric potential steps, special sets of exact solutions of
the relativistic wave equations with corresponding external fields are crucial
in this formulation. This technique was effectively used to describe the particle
creation effect in the Sauter field of the form $E(x)=E\cosh^{-2}\left(
x/L_{\mathrm{S}}\right)  $, in a constant electric field between two capacitor
plates, and in exponential time-independent electric steps, where the
corresponding exact solutions were available; see Refs.
\cite{x-case,L-field,x-exp}. However, there are only a few cases when such exact
solutions are known.

It was recently shown that near criticality, pair production exhibits
universal properties similar to those of continuous phase transitions
\cite{cr-regime1,cr-regime2}. In our terminology, this corresponds to the
situation when the Klein zone is relatively small. In the present article, we
show that there also exists a completely different type of universality,
believing that the Klein zone is quite extensive, so that the total number of
created pairs itself can be considered as a large parameter. Here, we develop a
nonperturbative approach that allows one to treat the vacuum instability
effects for arbitrary weakly inhomogeneous $x$-electric potential steps in the
absence of the corresponding exact solutions. The Schwinger effective action
method shows that when the probability of the pair creation is exponentially
small, there are certain relations between the probability and such total
physical quantities as a mean number density, a current density, and an
energy-momentum tensor of created pairs. However, for strong electric fields
such relations were established only for $t$-electric potential steps slowly
varying with time; see Ref. \cite{GavGit17}, and in a few existing exactly
solvable cases; see the review \cite{AdoGavGit17}. The approach developed by us in the
present article allows us to establish similar relations for arbitrary
weakly inhomogeneous $x$-electric potential steps.

The article is organized as follows. In Sec. \ref{S2}, we give a definition of
weakly inhomogeneous $x$-electric potential steps and present an overview of
the vacuum instability due to such backgrounds for existing exactly solvable
cases. We stress some universal features of the vacuum instability
in all these examples. In Sec. \ref{S3}, we present the approximation of a
weakly inhomogeneous field and derive universal forms for the flux density of
created particles and the probability of a vacuum to remain a vacuum. We
show that using these results, it is possible to calculate the physical
quantities for any slightly inhomogeneous but otherwise arbitrary constant
electric field. In this way, we reproduce the results obtained with the help of
existing exact solutions. Then we succeed to describe the vacuum instability
in electric fields where no exact solution of the corresponding Dirac equation
are known, namely, in the Gaussian peak $E(x)=E_{0}\exp\left[  -\left(
x/L_{\mathrm{G}}\right)  ^{2}\right]  $ and a special inverse square
field of a form\emph{ }$E(x)=E_{0}\left[  1+\left(  2x/L_{\mathrm{w}}\right)
^{2}\right]  ^{-1}$. Finally, in Sec.~\ref{S4}, we derive a
general expression for the current density vector of particles created from
a vacuum and relate it to the flux density of created particles obtained in the
approximation of a weakly inhomogeneous field.

\section{Weakly inhomogeneous potential steps: exactly solvable cases
\label{S2}}

Let us consider QED with a time-independent pure electric field\footnote{We recall that our system is placed in the ($d=D+1$)-dimensional Minkowski spacetime parametrized by the coordinates $X=\left(
X^{\mu},\ \mu=0,1,\ldots,D\right)  =\left(  t,\mathbf{r}\right)  $, $X^{0}=t$,
$\ \ \mathbf{r}=\left(  X^{1},\ldots,X^{D}\right)  $, $x=X^{1}$. It consists
of a Dirac field $\psi\left(  X\right)  $ interacting with an external
electromagnetic field $A^{\mu}(X)$ in the form of a $x$-electric potential
step.} $\mathbf{E}\left(  X\right)  =\mathbf{E}\left(  x\right)  =\left(
E\left(  x\right)  ,0,...,0\right)$. 
The inhomogeneous electric field $E\left(  x\right)$ has the form,\footnote{We use the system of units, where
$c=\hbar=1.$}%
\begin{align}
&  E\left(  x\right)  =E=\mathrm{const}>0,\ x\in S_{\mathrm{int}}=\left(
x_{\mathrm{L}},x_{\mathrm{R}}\right)  ;\ \nonumber\\
&  E\left(  x\right)  =0,\ x\in S_{\mathrm{L}}=\left(  -\infty,x_{\mathrm{L}%
}\right]  ,\ x\in S_{\mathrm{R}}=\left[  x_{\mathrm{R}},\infty\right)  .
\label{sc0}%
\end{align}
We assume that the basic Dirac particle is an electron with the mass $m$ and
the charge $-e$, $e>0$, and the positron is its antiparticle. The electric
field under consideration accelerates the electrons along the $x$ axis in the
negative direction and the positrons along the $x$ axis in the positive
direction. Potentials of the corresponding electromagnetic field $A^{\mu
}\left(  X\right)  $ can be chosen as%
\begin{equation}
A^{\mu}\left(  X\right)  =\left(  A^{0}\left(  x\right)  ,A^{j}%
=0,\ j=1,2,\ldots,D\right)  , \label{2.3}%
\end{equation}
so that $E\left(  x\right)  =-\partial_{x}A_{0}\left(  x\right)  $.

We call the electric field $E(x)$ a weakly inhomogeneous electric field on a
spatial interval $\Delta l$ if the following condition holds true:%
\begin{equation}
\left\vert \frac{\overline{\partial_{x}E(x)}\Delta l}{\overline{E(x)}%
}\right\vert \ll1,\ \ \Delta l/\Delta l_{\mathrm{st}}^{\mathrm{m}}\gg1,
\label{sc.1}%
\end{equation}
where $\overline{E(x)}$ and $\overline{\partial_{x}E(x)}$ are the mean values of
$E(x)$ and $\partial_{x}E(x)$ on the spatial interval $\Delta l$,
respectively, and $\Delta l$ is significantly larger than the length scale
$\Delta l_{\mathrm{st}}^{\mathrm{m}}$, which is%
\begin{equation}
\Delta l_{\mathrm{st}}^{\mathrm{m}}=\Delta l_{\mathrm{st}}\max\left\{
1,m^{2}/e\overline{E(x)}\right\}  ,\ \ \Delta l_{\text{\textrm{st}}}=\left[
e\overline{E(x)}\right]  ^{-1/2}. \label{sc.2}%
\end{equation}

Note that the length scale $\Delta l_{\mathrm{st}}^{\mathrm{m}}$ appears in
Eq.~(\ref{sc.1}) as the length scale when the perturbation theory with respect
to the electric field breaks down and the Schwinger (nonperturbative)
mechanism is primarily responsible for the pair creation. In what follows, we
show that this condition is sufficient. We are primarily interested
in strong electric fields, $m^{2}/e\overline{E(x)}\lesssim1$. In this case,
the second inequality in Eq.~(\ref{sc.1}) is simplified to the form $\Delta
l/\Delta l_{\mathrm{st}}\gg1$, in which the mass $m$ is absent. In such cases,
the potential of the corresponding electric step hardly differs from the
potential of a uniform electric field,%
\begin{equation}
U(x)=-eA_{0}(x)\approx U_{\mathrm{const}}(x)=e\overline{E(x)}x+U_{0},
\label{sc.3}%
\end{equation}
on the interval $\Delta l$, where $U_{0}$ is a given constant. We see this
behavior for the fields of known exact solvable cases for $x$-electric
potential steps, namely, the Sauter field, the $L$-constant electric field,
and the exponential peak field.

The magnitude of the corresponding $x$-potential step is%
\begin{equation}
\Delta U=U_{\mathrm{R}}-U_{\mathrm{L}}>0,\ \ U_{\mathrm{R}}=-eA_{0}%
(+\infty),\ \ U_{\mathrm{L}}=-eA_{0}(-\infty). \label{f.7}%
\end{equation}
We are interested in electron-positron pair creation that exists for the critical
steps, $\Delta U>2m$; see Ref. \cite{x-case} for details. The Dirac equation
with an $x$-electric potential step has the form,
\begin{align}
&  i\partial_{0}\psi(X)=\hat{H}\psi(X),\ \ \hat{H}=\gamma^{0}\left(
-i\gamma^{j}\partial_{j}+m\right)  +U(x),\ \ j=1,\ldots D,\nonumber\\
&  \psi(X)=\exp\left(  -ip_{0}t+i\mathbf{p}_{\perp}\mathbf{r}_{\perp}\right)
\psi_{n}(x),\ \ \mathbf{p}_{\perp}=\left(  p^{2},\ldots,p^{D}\right)
,\nonumber\\
&  \psi_{n}(x)=\left\{  \gamma^{0}\left[  p_{0}-U(x)\right]  -\gamma^{1}%
\hat{p}_{x}-\mathbf{\gamma}_{\perp}\mathbf{p}_{\perp}+m\right\}  \varphi
_{n}^{(\chi)}(x)\upsilon_{\chi},\ \nonumber\\
&  \gamma^{0}\gamma^{1}\upsilon_{\chi}=\chi\upsilon_{\chi},\ \ \chi
=\pm1,\ \ \upsilon_{\chi^{\prime},\sigma^{\prime}}^{\dag}\upsilon_{\chi
,\sigma}=\delta_{\chi^{\prime}\chi}\delta_{\sigma^{\prime}\sigma}.
\label{ap.2}%
\end{align}
Here $\psi(x)$ is a $2^{[d/2]}$-component spinor, $[d/2]$ stands for the
integer part of $d/2$, $m\neq0$ is the electron mass, and $\gamma^{\mu}$ are the
$\gamma$ matrices in $d$ dimensions. The complete set of solutions of such a
Dirac equation is determined by the functions $_{\zeta}\varphi_{n}(x)$ and
$^{\zeta}\varphi_{n}(x)$ with special right and left asymptotics ($\zeta=\pm$)
at $x\in S_{\mathrm{L}}$ and $x\in S_{\mathrm{R}}$, respectively. These
solutions are parametrized by the set of quantum numbers $n=(p_{0}%
,\mathbf{p}_{\bot},\sigma)$ where $p_{0}$ stands for total energy,
$\mathbf{p}_{\bot}$ is transversal momentum (the index $\perp$ stands for
components of momentum that are perpendicular to the electric field), and
$\sigma$ is spin polarization. The solutions $\varphi_n^{(\chi)}(x)$ which only differ by the values of $\chi$, are linearly dependent. Because of this, it suffices to work with solutions corresponding to one of the possible two values of $\chi$; so here and in what follows we omit the subscript $\chi$ in these solutions, implying that the spin quantum number $\chi$ is fixed in a certain way.

A critical step produces electron-positron pairs in the Klein zone $\Omega_{3}$,
defined by the double inequality,
\begin{equation}
\Omega_{3}:\ \ U_{\mathrm{L}}+\pi_{\perp}\leq p_{0}\leq U_{\mathrm{R}}%
-\pi_{\perp},\ \ 2\pi_{\perp}\leq\Delta U, \label{Klein}%
\end{equation}
where $\pi_{\bot}=\sqrt{\mathbf{p}_{\bot}^{2}+m^{2}}$. In this range, initial
states are determined by functions $_{-}\varphi_{n}$ for a positron
and$\ ^{-}\varphi_{n}$ for an electron while final states are determined by
functions $_{+}\varphi_{n}$ for a positron and$\ ^{+}\varphi_{n}$ for an electron
(see Ref. \cite{x-case} for details). The latter functions satisfy the
following asymptotic conditions:
\begin{align}
&  _{\;\zeta}\varphi_{n}\left(  x\right)  =\ _{\zeta}\mathcal{N}\exp\left[
ip^{\mathrm{L}}\left(  x-x_{\mathrm{L}}\right)  \right]  ,\ \ x\in
S_{\mathrm{L}},\nonumber\\
&  ^{\;\zeta}\varphi_{n}\left(  x\right)  =\ ^{\zeta}\mathcal{N}\exp\left[
ip^{\mathrm{R}}\left(  x-x_{\mathrm{R}}\right)  \right]  ,\ \ x\in
S_{\mathrm{R}},\nonumber\\
&  p^{\mathrm{L}}=\zeta\sqrt{\left[  \pi_{0}\left(  \mathrm{L}\right)
\right]  ^{2}-\pi_{\bot}^{2}},\ \ p^{\mathrm{R}}=\zeta\sqrt{\left[  \pi
_{0}\left(  \mathrm{R}\right)  \right]  ^{2}-\pi_{\bot}^{2}},\ \zeta
=\pm\ ,\nonumber\\
&  \pi_{0}\left(  \mathrm{L/R}\right)  =p_{0}-U_{\mathrm{L/R}}. \label{L3}%
\end{align}
The constants $^{\zeta}\mathcal{N}$ and $_{\zeta}\mathcal{N}$ are
normalization factors with respect to the inner product on the $x$-constant
hyperplane \cite{x-case}, %
\begin{align}
&  \left(  \ _{\zeta}\psi_{n},\ _{\zeta^{\prime}}\psi_{n^{\prime}}\right)
_{x}=\zeta\eta_{\mathrm{L}}\delta_{\zeta,\zeta^{\prime}}\delta_{n,n^{\prime}%
},\ \ \eta_{\mathrm{L}}=\mathrm{sgn\ }\pi_{0}\left(  \mathrm{L}\right)
,\nonumber\\
&  \left(  \ ^{\zeta}\psi_{n},\ ^{\zeta^{\prime}}\psi_{n^{\prime}}\right)
_{x}=\zeta\eta_{\mathrm{R}}\delta_{\zeta,\zeta^{\prime}}\delta_{n,n^{\prime}%
},\ \ \eta_{\mathrm{R}}=\mathrm{sgn\ }\pi_{0}\left(  \mathrm{R}\right)
,\nonumber\\
&  \left(  \psi,\psi^{\prime}\right)  _{x}=\int\psi^{\dag}\left(  X\right)
\gamma^{0}\gamma^{1}\psi^{\prime}\left(  X\right)  dtd\mathbf{r}_{\bot}\ ,
\label{c3}%
\end{align}
having the form
\begin{align}
&  ^{\zeta}\mathcal{N}=\ ^{\zeta}CY,\ _{\zeta}\mathcal{N}=\ _{\zeta
}CY,\ Y=(V_{\perp}T)^{-1/2},\nonumber\\
&  ^{\zeta}C=\left[  2\left\vert p^{\mathrm{R}}\right\vert \left\vert \pi
_{0}(\mathrm{R})-\chi p^{\mathrm{R}}\right\vert \right]  ^{-1/2},\ _{\zeta
}C=\left[  2\left\vert p^{\mathrm{L}}\right\vert \left\vert \pi_{0}%
(\mathrm{L})-\chi p^{\mathrm{L}}\right\vert \right]  ^{-1/2}, \label{L4}%
\end{align}
where $V_{\bot}$ is the spatial volume of the $(d-1)$-dimensional hypersurface
orthogonal to the electric field direction $x$, and $T$ is the time duration
of the electric field. Both $V_{\bot}$ and $T$ are macroscopically large. The
functions $^{\;\zeta}\varphi_{n}\left(  x\right)  $ and $_{\;\zeta}\varphi
_{n}\left(  x\right)  $ are connected by the decomposition%
\begin{align}
^{\;\zeta}\varphi_{n}\left(  x\right)  =\ _{\;+}\varphi_{n}\left(  x\right)
g\left(  _{+}\left\vert ^{\zeta}\right.  \right)  -\ _{\;-}\varphi_{n}\left(
x\right)  g\left(  _{-}\left\vert ^{\zeta}\right.  \right), \ \ \left(  \ _{\pm}\psi_{n},\ ^{\zeta}\psi_{n^{\prime}}\right)_{x}=
\delta_{n,n^{\prime}}g\left(  _{\pm}\left\vert ^{\zeta}\right.  \right)
\label{dec}%
\end{align}
in the Klein zone.

Partial vacuum states for a given $n\notin\Omega_{3}$ are stable. The
differential mean numbers of electrons and positrons from the electron-positron
pairs created are nonzero for $n\in\Omega_{3}$ only and are defined as the average values,
\begin{align}
&  N_{n}^{a}\left(  \mathrm{out}\right)  =\left\langle 0,\mathrm{in}\left\vert
\ ^{+}a_{n}^{\dagger}(\mathrm{out})\ ^{+}a_{n}(\mathrm{out})\right\vert
0,\mathrm{in}\right\rangle ,\nonumber\\
&  N_{n}^{b}\left(  \mathrm{out}\right)  =\left\langle 0,\mathrm{in}\left\vert
\ _{+}b_{n}^{\dagger}(\mathrm{out})\ _{+}b_{n}(\mathrm{out})\right\vert
0,\mathrm{in}\right\rangle ,\ \label{Nab}%
\end{align}
where\ $^{+}a_{n}^{\dagger}(\mathrm{out})$ and $\ ^{+}a_{n}(\mathrm{out})$ are
the creation and annihilation operators of final electrons while $_{+}%
b_{n}^{\dagger}(\mathrm{out})$ and$\ _{+}b_{n}(\mathrm{out})$ are the creation
and annihilation operators of final positrons, respectively. We define two
vacuum vectors $\left\vert 0,\mathrm{in}\right\rangle $ and $\left\vert
0,\mathrm{out}\right\rangle $; the first of which is the\ vacuum vector for
all annihilation operators of initial particles, and the other is the vacuum
vector for all annihilation operators of final particles. The numbers
$N_{n}^{a}\left(  \mathrm{out}\right)  $ and $N_{n}^{b}\left(  \mathrm{out}%
\right)  $ are equal and represent the number of pairs created, $N_{n}%
^{\mathrm{cr}}$, which can be expressed as
\begin{equation}
N_{n}^{b}\left(  \mathrm{out}\right)  =N_{n}^{a}\left(  \mathrm{out}\right)
=N_{n}^{\mathrm{cr}}=|g\left(  _{+}|^{-}\right)  |^{-2},\ \ n\in\Omega_{3}.
\label{meanN}%
\end{equation}

The exact solutions of the Dirac equation are known for the following inhomogeneous electric fields.

\textrm{(i)} The Sauter electric field and its vector potential have the form%
\begin{equation}
E(x)=E_{0}\cosh^{-2}(x/L_{\mathrm{S}}),\ \ A_{0}(x)=-L_{\mathrm{S}}E_{0}%
\tanh(x/L_{\mathrm{S}}),\ \ L_{\mathrm{S}}>0, \label{sc.5}%
\end{equation}
where the parameter $L_{\mathrm{S}}$ sets the scale. The corresponding
solutions of the Dirac equation\ $_{\zeta}\varphi_{n}(x)$ and $^{\zeta}%
\varphi_{n}(x)$ and the number of particles created $N_{n}^{\mathrm{cr}}$ for
this potential were found in Ref. \cite{x-case}. This field is considered a
weakly inhomogeneous one if the following condition holds true:
\begin{equation}
eE_{0}L_{\mathrm{S}}^{2}\gg\max\left\{  1,m^{2}/eE_{0}\right\}  . \label{sc.6}%
\end{equation}
The leading contribution to the total number of pairs created is formed in the
inner part of the Klein zone $\Omega_{3}$,
\begin{equation}
\left\vert p_{0}\right\vert <eE_{0}L_{\mathrm{S}}-K/L_{\mathrm{S}}%
,\ \ \pi_{\bot}<K_{\bot}/L_{\mathrm{S}},\ \ \pi_{\bot}^{2}=m^{2}%
+\mathbf{p}_{\bot}^{2}, \label{sc.7}%
\end{equation}
where $K$ is
\begin{equation}
K=L_{\mathrm{S}}\sqrt{(km)^{2}+\pi_{\bot}^{2}},\ \label{sc.8}%
\end{equation}
while $K_{\bot}$ and $k\gtrsim1$ are given arbitrary numbers obeying the
inequalities%
\begin{equation}
km\ll eE_{0}L_{\mathrm{S}},\ \ mL_{\mathrm{S}}\ll K_{\bot}\ll eEL_{\mathrm{S}%
}^{2}. \label{sc.8a}%
\end{equation}
The differential mean number of particles created in this range is
\cite{x-case}
\begin{equation}
N_{n}^{\mathrm{cr}}\approx N_{n}^{\mathrm{as}}=e^{-\pi\tau},\ \,\tau
=\exp\left[  -\pi L_{\mathrm{S}}\left(  2eE_{0}L_{\mathrm{S}}-\left\vert
p^{\mathrm{R}}\right\vert -\left\vert p^{\mathrm{L}}\right\vert \right)
\right]  , \label{sc.9}%
\end{equation}
where left and right asymptotic momenta $\left\vert p^{\mathrm{R}}\right\vert
$ and $\left\vert p^{\mathrm{L}}\right\vert $ are defined by Eq. (\ref{L3}).
This distribution has its maximum at zero energy, $p_{0}=0$, which coincides
with the number of particles created by a uniform electric field,
\begin{equation}
N_{n}^{0}=\left.  N_{n}^{\mathrm{cr}}\right\vert _{p_{0}=0}=\exp\left[
-\pi\lambda\right]  ,\ \ \lambda=\pi_{\bot}^{2}/eE_{0}. \label{sc.10}%
\end{equation}

\textrm{(ii)} The so-called $L$-constant field does not change within
the spatial region $L$ and is zero outside of it,%
\begin{equation}
E(x)=\left\{
\begin{array}
[c]{l}%
0,\ \ x\in S_{\mathrm{L}}\\
E_{0}>0,\ \ x\in S_{\mathrm{int}}\\
0,\ \ x\in S_{\mathrm{R}}%
\end{array}
\right.  \Longrightarrow A_{0}(x)=\left\{
\begin{array}
[c]{l}%
-E_{0}x_{\mathrm{L}},\ \ x\in S_{\mathrm{L}}\\
-E_{0}x,\ \ x\in S_{\mathrm{int}}\\
-E_{0}x_{\mathrm{R}},\ \ x\in S_{\mathrm{R}}%
\end{array}
\right.  , \label{sc.11}%
\end{equation}
where the regions are $S_{\mathrm{L}}=\left(  -\infty,x_{\mathrm{L}}\right]
$,$\ S_{\mathrm{R}}=\left[  x_{\mathrm{R}},+\infty\right)  $, $S_{\mathrm{int}%
}=\left(  x_{\mathrm{L}},x_{\mathrm{R}}\right)  ,$ and we chose that
$x_{\mathrm{L}}=-L/2$,$\ x_{\mathrm{R}}=L/2$. The corresponding solutions of
the Dirac equation\ $_{\zeta}\varphi_{n}(x)$ and $^{\zeta}\varphi_{n}(x)$ were
found in Ref. \cite{L-field}. The $L$-constant field can be considered a
weakly inhomogeneous field if%
\begin{equation}
\sqrt{eE_{0}}L\gg\max\left(  1,m^{2}/eE_{0}\right)  . \label{sc.12}%
\end{equation}
The leading contribution to the number of particles created is formed in the inner part $D$ of the Klein zone $\Omega_{3}$:%
\begin{align}
&  \sqrt{\lambda}<K_{\bot},\ \ \left\vert p_{0}\right\vert /\sqrt
{eE_{0}}\leq\sqrt{eE_{0}}L/2-K,\nonumber\\
&  \sqrt{eE_{0}}L/2\gg K\gg K_{\bot}^{2}\gg\max\left\{  1,m^{2}/eE_{0}%
\right\}  . \label{sc.12a}%
\end{align}
The leading contribution to $N_{n}^{\mathrm{cr}}$ has the form (\ref{sc.10}).

\textrm{(iii)} An exponential peak electric field has the following structure.
Its first part is increasing exponentially on the spatial interval $I=\left(
-\infty,0\right]  $ and reaches its maximum value $E_{0}$ at $x=0$. The other
part decreases exponentially from the same value $E_{0}$ on the spatial
interval $II=\left(  0,+\infty\right)  $. The potential $A_{0}(x)$ and the
electric field $E(x)$ have the form%
\begin{equation}
E(x)=E_{0}\left\{
\begin{array}
[c]{l}%
e^{k_{1}x},\ \ x\in I\\
e^{-k_{2}x},\ \ x\in II
\end{array}
\right.  \Longrightarrow A_{0}(x)=E_{0}\left\{
\begin{array}
[c]{l}%
k_{1}^{-1}\left(  -e^{k_{1}x}+1\right)  ,\ \ x\in I\\
k_{2}^{-1}\left(  e^{-k_{2}x}-1\right)  ,\ \ x\in II
\end{array}
\right.  , \label{sc.13}%
\end{equation}
where $k_{1}$ and $k_{2}$ are some positive constants. The corresponding
solutions of the Dirac equation $_{\zeta}\varphi_{n}(x)$ and $^{\zeta}%
\varphi_{n}(x)$ were found in Ref. \cite{x-exp}.

The case of a weakly inhomogeneous exponential peak corresponds to small
values of $k_{1}$ and $k_{2}$, and is characterized by the condition,
\begin{equation}
\min(h_{1},h_{2})\gg\max\left(  1,m^{2}/eE_{0}\right)  ,\ \ h_{1,2}%
=2eE_{0}/k_{1,2}^{2}. \label{sc.16}%
\end{equation}
The main contributions to the number of particles created $N_{n}^{\mathrm{cr}%
}$ are formed in the ranges of quantum numbers $\pi_{\perp}<\pi_{0}%
(\mathrm{L})\leq eE_{0}k_{1}^{-1}$ and $-eE_{0}k_{2}^{-1}\geq\pi
_{0}(\mathrm{R})>-\pi_{\perp}$, and have the following forms:%
\begin{subequations}
\begin{align}
&  N_{n}^{\mathrm{cr}}\approx\exp\left[  -\frac{2\pi}{k_{1}}\left(  \pi
_{0}(\mathrm{L})-\left\vert p^{\mathrm{L}}\right\vert \right)  \right]
,\ \ \pi_{\perp}<\pi_{0}(\mathrm{L})\leq eE_{0}k_{1}^{-1},\label{sc.17a}\\
&  N_{n}^{\mathrm{cr}}\approx\exp\left[  -\frac{2\pi}{k_{2}}\left(  \left\vert
\pi_{0}(\mathrm{R})\right\vert -\left\vert p^{\mathrm{R}}\right\vert \right)
\right]  ,\ \ -eE_{0}k_{2}^{-1}\geq\pi_{0}(\mathrm{R})>-\pi_{\perp}.
\label{sc.17}%
\end{align}
\end{subequations}

In the examples under discussion, the intervals of growth and decay are
described by nearly the same functional form; that is, increasing and
decreasing components of the fields are almost symmetric. One can consider a
strongly asymmetric configuration of the peak field, when one of the
parameters $k$, for example, $k_{1}$, is sufficiently large, so that%
\begin{equation}
eE_{0}k_{1}^{-2}\ll1,\ \ \left\vert p^{\mathrm{L}}\right\vert /k_{1}\ll1,
\label{sc.18}%
\end{equation}
while the second one, $k_{2}>0$, is arbitrary. This field is a weakly
inhomogeneous one if
\begin{equation}
h_{2}\gg\max\left(  1,m^{2}/eE_{0}\right)  . \label{sc.19}%
\end{equation}
The leading contribution to the number of particles created is formed in the
range of quantum numbers $-eE_{0}k_{2}^{-1}\geq\pi_{0}(\mathrm{R})>-\pi
_{\perp},$ and coincides with the Eq. (\ref{sc.17}). Note that
this situation can be easily transformed to the case with a large $k_{2}$ and
arbitrary $k_{1}$ by a simultaneous change $k_{1}\leftrightarrows k_{2}$ and
$\pi_{0}(\mathrm{L})\leftrightarrows-\pi_{0}(\mathrm{R})$.

For weakly inhomogeneous electric fields, the differential mean numbers of
electron-positron pairs created from the vacuum are almost constant over the
wide range of energies $p_{0}$ for any given transversal momenta
$\mathbf{p}_{\bot}$, even if these distributions are different for different
fields. Furthermore, for all exactly solvable cases, there are wide subranges
where the distributions $N_{n}^{\mathrm{cr}}$ coincide with the corresponding
distributions $N_{n}^{0}$ in a constant uniform electric field, given by Eq.
(\ref{sc.10}). We call this phenomenon the stabilization of the particle
creation effect. In these subranges of quantum numbers, $N_{n}^{\mathrm{cr}}$
hardly depend on the details of how the field grows and decays. We note that
the similar effect takes place for slowly varying electric fields; see Ref.
\cite{GavGit17} for the details.

The total number of pairs $N^{\mathrm{cr}}$ created from vacuum by an
$x$-electric potential step can be calculated by the summation over all possible
quantum numbers in the Klein zone $\Omega_{3}$,%
\begin{equation}
N^{\mathrm{cr}}=\sum_{n\in\Omega_{3}}N_{n}^{\mathrm{cr}}. \label{Ncr}%
\end{equation}
\emph{ }It is proportional to the so-called transversal space-time volume
$V_{\perp}T$, $N^{\mathrm{cr}}=V_{\perp}Tn^{\mathrm{cr}}$, where $d$ labels
the space-time dimensions, and the corresponding densities $n^{\mathrm{cr}}$
have the form%
\begin{equation}
n^{\mathrm{cr}}=\frac{J_{(d)}}{(2\pi)^{d-1}}\int_{\Omega_{3}}dp_{0}%
d\mathbf{p}_{\bot}N_{n}^{\mathrm{cr}}. \label{sc.20}%
\end{equation}
In fact, $n^{\mathrm{cr}}$ is the total flux density of created particles. In
the latter expression the summations over the energies and transversal momenta
were transformed into integrals, and the summation over spin projections was
fulfilled, $J_{(d)}=2^{\left[  d/2\right]  -1}$ (the square brackets mean the
integer part of $d/2$). In weakly inhomogeneous fields, the magnitude of a
potential step\emph{ }$\Delta U$ is large and can be used as a large
parameter. The integral on the right-hand side can be approximated by an
integral over a subrange $D$ that gives the dominant contribution with respect
to the total increment to the flux density of created particles,%
\begin{equation}
n^{\mathrm{cr}}\approx\tilde{n}^{\mathrm{cr}}=\frac{J_{(d)}}%
{(2\pi)^{d-1}}\int_{D}dp_{0}d\mathbf{p}_{\bot}N_{n}^{\mathrm{cr}}.
\label{sc.21}%
\end{equation}
The exact form of the subrange $D$ for each particular field must be
determined separately. The dominant contributions $\tilde{n}^{\mathrm{cr}}$
are proportional to the magnitude of potential steps (and then the maximum
increments of a particle energy), which, in general, differs for different
fields and, for example, has the following forms in the exactly solvable
cases $\mathrm{(i)}$, $\mathrm{(ii)}$, and $\mathrm{(iii)}$:%
\begin{align}
&  \mathrm{(i)}\text{\ }\Delta U_{S}=2eE_{0}L_{\mathrm{S}}\text{ \textrm{for
the Sauter field}},\nonumber\\
&  \mathrm{(ii)}\text{\ }\Delta U_{L}=eE_{0}L\text{ }\mathrm{for\ the}%
L\mathrm{-constant\ field},\nonumber\\
&  \mathrm{(iii)}\text{\ }\Delta U_{P}=eE_{0}\left(  k_{1}^{-1}+k_{2}%
^{-1}\right)  \ \mathrm{for\ the\ peak\ field}. \label{sc.22}%
\end{align}
We note that $\Delta U_{P}$ corresponds to the case of a strongly asymmetric
exponential field configuration at $k_{1}^{-1}\rightarrow0$\textrm{ }(or at
$k_{2}^{-1}\rightarrow0$).

In terms of the introduced quantities (\ref{sc.22}), the densities $\tilde
{n}^{\mathrm{cr}}$ in the exactly solvable cases under consideration have the forms \cite{x-case,L-field,x-exp},
\begin{align}
&  \mathrm{(i)}\text{\ }\tilde{n}^{\mathrm{cr}}=r^{\mathrm{cr}}\frac{\Delta
U_{S}}{2eE_{0}}\delta\ \text{\textrm{for the Sauter field}},\nonumber\\
&  \mathrm{(ii)}\text{\ }\tilde{n}^{\mathrm{cr}}=r^{\mathrm{cr}}\frac{\Delta
U_{L}}{eE_{0}}\ \mathrm{for\ the\ }L\mathrm{-constant\ field},\nonumber\\
&  \mathrm{(iii)}\ \tilde{n}^{\mathrm{cr}}=r^{\mathrm{cr}}\frac{\Delta U_{P}%
}{eE_{0}}G\left(  \frac{d}{2},\pi\frac{m^{2}}{eE_{0}}\right)
\ \mathrm{for\ the\ peak\ field}, \label{sc.23}%
\end{align}
where%
\begin{align}
&  r^{\mathrm{cr}}=\frac{J_{(d)}\left(  eE_{0}\right)  ^{d/2}}{(2\pi)^{d-1}%
}\exp\left\{  -\pi\frac{m^{2}}{eE_{0}}\right\}  ,\ G\left(  \alpha,x\right)
=\int_{1}^{\infty}\frac{ds}{s^{\alpha+1}}e^{-x(s-1)}=e^{x}x^{\alpha}%
\Gamma(-\alpha,x),\nonumber\\
&  \delta=\int_{0}^{\infty}dt\ t^{-1/2}\left(  t+1\right)  ^{-(d+1)/2}%
\exp\left(  -t\pi\frac{m^{2}}{eE_{0}}\right)  =\sqrt{\pi}\Psi\left(  \frac
{1}{2},\frac{2-d}{2};\pi\frac{m^{2}}{eE_{0}}\right)  . \label{sc.24}%
\end{align}
Here, the $\Gamma(-\alpha,x)$ is the incomplete gamma function and $\Psi\left(
a,b;x\right)  $ is the confluent hypergeometric function\footnote{In Ref.
\cite{x-case} the result for a Sauter field was obtained under an unnecessary
assumption that $\lambda>1$. However, the final form of $\delta=\sqrt{\pi}%
\Psi\left(  \frac{1}{2},\frac{2-d}{2};\pi\frac{m^{2}}{eE}\right)  $ is given
correctly for an arbitrary $m^{2}/eE$ in Ref. \cite{x-case}. We have to correct
its derivation as follows.\ Note that\textrm{ }for any $\lambda$ one can see
that $N_{n}^{\mathrm{as}}$, given by Eq.~(\ref{sc.9}), is exponentially small
if $km\sim\pi_{\bot}$, where $k$ is any given number satisfying inequality
$k\ll\pi mL_{\mathrm{S}}/2$. Therefore, the range $\pi_{\bot}\ll km$ is of
interest. In this range, the approximation $\tau\approx
eEL_{\mathrm{S}}^{2}\pi_{\perp}^{2}\left[  (eEL_{\mathrm{S}})^{2}-p_{0}%
^{2}\right]  ^{-1}$ holds true. Taking into account the relation between $\tau$ and
$p_{0}$, one can find a correct form of $\delta$ that is presented in
Eq.~(\ref{sc.24}).}.

Equating the densities $\tilde{n}^{\mathrm{cr}}$ for a Sauter-like field
$\mathrm{(i)}$ and for the peak field $\mathrm{(iii)}$ to the density
$n^{\mathrm{cr}}$ for the $L$-constant field, we find effective lengths
$L_{\mathrm{eff}}$ of the interval of the field action for both cases,
\begin{align}
&  \mathrm{(i)}\text{\ }L_{\mathrm{eff}}=L_{\mathrm{S}}\delta,\nonumber\\
&  \mathrm{(iii)}\text{ }L_{\mathrm{eff}}=\left(  k_{1}^{-1}+k_{2}%
^{-1}\right)  G\left(  \frac{d}{2},\pi\frac{m^{2}}{eE_{0}}\right)  .
\label{sc.25}%
\end{align}
Note that the effective length $L_{\mathrm{eff}}$ for a strongly asymmetric
exponential field configuration is given by the second line in Eq.
(\ref{sc.25}) as $k_{1}^{-1}\rightarrow0$ (or $k_{2}^{-1}\rightarrow0$). It is
obvious that $L_{\mathrm{eff}}=L$ for the $L$-constant field. One can say that
the Sauter field, the peak electric field, and the asymmetric exponential
field with the same $L_{\mathrm{eff}}$ are equivalent to the $L$-constant
field with respect to the pair production. Note that the factors $G$ and
$\delta$ in Eq. (\ref{sc.24}) for weak $\left(  m^{2}/eE_{0}\gg1\right)  $ and
strong $\left(  m^{2}/eE_{0}\ll1\right)  $ electric fields can be approximated
as%
\begin{align}
G\left(  \frac{d}{2},\pi\frac{m^{2}}{eE_{0}}\right)   &  \approx\frac{eE_{0}%
}{\pi m^{2}},\ \delta\approx\frac{\sqrt{eE_{0}}}{m},\text{ }\frac{m^{2}%
}{eE_{0}}\gg1;\nonumber\\
G\left(  \frac{d}{2},\pi\frac{m^{2}}{eE_{0}}\right)   &  \approx\frac{2}%
{d},\ \delta\approx\frac{\sqrt{\pi}\Gamma(d/2)}{\Gamma(d/2+1/2)},\ \frac
{m^{2}}{eE_{0}}\ll1. \label{sc.26}%
\end{align}
One can compare the scale lengths for the cases \textrm{(i)}, \textrm{(ii)},
and \textrm{(iii)}, given by Eqs. (\ref{sc.6}), (\ref{sc.12}), and
(\ref{sc.18}), for the same $E_{0}$ and energy increments $\Delta
U_{S}=\Delta U_{L}=\Delta U_{P}$ (in this case, $2L_{\mathrm{S}}=L=k_{1}%
^{-1}+k_{2}^{-1}$). The condition Eq. (\ref{sc.12}) is stronger than Eqs.
(\ref{sc.6}) and (\ref{sc.18}) if the fields are weak, whereas they are equivalent
if the fields are strong. For this reason, defining the scale $\Delta
l_{\mathrm{st}}^{\mathrm{m}}$ in general terms, we choose the form (\ref{sc.2}).

Initial $\left\vert 0,\mathrm{in}\right\rangle $ and final $\left\vert
0,\mathrm{out}\right\rangle $ vacua do not coincide. In the general case, the
probability for a vacuum to remain a vacuum $P_{\mathrm{v}}=\left\vert
\langle0,\mathrm{out}|0,\mathrm{in}\rangle\right\vert ^{2}$ can be expressed
via the distribution $N_{n}^{\mathrm{cr}}$ as
\begin{equation}
P_{\mathrm{v}}=\prod_{n\in\Omega_{3}}\left(  1-N_{n}^{\mathrm{cr}}\right)  ,
\label{Pv}%
\end{equation}
whereas, for weakly inhomogeneous electric fields, it can be written%
\begin{equation}
P_{\mathrm{v}}\approx\prod_{n\in D}\left(  1-N_{n}^{\mathrm{cr}}\right)  ,
\label{Papp}%
\end{equation}
where $D$ is the same subrange that gives the dominant contribution to
the flux density of created particles (\ref{sc.21}). The probability
$P_{\mathrm{v}}$ is given by similar forms for the Sauter field $\mathrm{(i)}%
$, the $L$-constant field $\mathrm{(ii)}$, and the peak field $\mathrm{(iii)}%
$, respectively, with the corresponding $N^{\mathrm{cr}}$,
\begin{align}
&  P_{\mathrm{v}}=\exp\left(  -\mu V_{\perp}T\tilde{n}^{\mathrm{cr}}\right)
,\ \mu=\sum_{l=0}^{\infty}\frac{\epsilon_{l+1}}{\left(  l+1\right)  ^{d/2}%
}\exp\left(  -l\pi\frac{m^{2}}{eE_{0}}\right)  ,\nonumber\\
&  \mathrm{(i)}\ \ \epsilon_{l}=\epsilon_{l}^{S}=\delta^{-1}\sqrt{\pi}%
\Psi\left(  \frac{1}{2},\frac{2-d}{2};l\pi\frac{m^{2}}{eE_{0}}\right)
,\nonumber\\
&  \mathrm{(ii)}\ \ \epsilon_{l}=\epsilon_{l}^{L}=1,\nonumber\\
&  \mathrm{(iii)}\ \ \epsilon_{l}=\epsilon_{l}^{P}=G\left(  \frac{d}{2}%
,l\pi\frac{m^{2}}{eE_{0}}\right)  \left[  G\left(  \frac{d}{2},\pi\frac{m^{2}%
}{eE_{0}}\right)  \right]  ^{-1}. \label{sc.27}%
\end{align}
In the case of a weak field ($m^{2}/eE_{0}\gg1$), $\epsilon_{l}^{S}\approx
l^{-1/2}$ for the Sauter field, $\epsilon_{l}^{P}\approx l^{-1}$ for the peak
field, and $\exp(-\pi m^{2}/eE_{0})\ll1$. Then $\mu\approx1$ for all the cases
in Eq. (\ref{sc.27}) and we have a universal relation $N^{\mathrm{cr}}%
\approx\ln P_{\mathrm{v}}^{-1}$. In the case of a strong field ($m^{2}%
/eE_{0}\ll1$), all the terms with the different $\epsilon_{l}^{S}$ and
$\epsilon_{l}^{P}$ contribute significantly to the sum in Eq. (\ref{sc.27}), if
$l\pi m^{2}/eE_{0}\lesssim1$, and the quantities $\mu$ for the Sauter and peak
fields differ essentially from the case of the $L$-constant field. Consequently,
in this situation, one cannot derive a universal relation between
$P_{\mathrm{v}}$ and $\tilde{n}^{\mathrm{cr}}$ from particular cases given by
Eq. (\ref{sc.27}). In addition, it should be noted that in the case of a
strong field, when known semiclassical approaches are not applicable, the
probability $P_{\mathrm{v}}$ (unlike the total number $N^{\mathrm{cr}}$) no
longer has a direct relation to the vacuum mean values of the physical quantities
discussed above. Therefore, to study a universal behavior of the vacuum
instability in weakly inhomogeneous fields one should derive first a universal
form for the total flux density $\tilde{n}^{\mathrm{cr}}$.

\section{Universal behavior of the flux density of created pairs in a strong
electric field \label{S3}}

Unlike the case of uniform time-dependent electric fields, in constant
inhomogeneous electric fields, there is a critical surface in space of
particle momenta, which separates the Klein zone $\Omega_{3}$, defined by
inequality (\ref{Klein}), from the adjacent ranges $\Omega_{2}$ and $\Omega_{4}$
(we use notation defined in Ref. \cite{x-case}). In the ranges $\Omega_{2}$ and
$\Omega_{4}$, the work of an electric field is sufficient to ensure the total
reflection for electrons and positrons, respectively, but is not sufficient to
produce pairs from vacuum. Accordingly, it is expected that for any
nonpathological field configuration, the pair creation vanishes close to this
critical surface, so that $N_{n}^{\mathrm{cr}}\rightarrow0$ if $n$ tends to
the boundary with either the range $\Omega_{2}$ ($\left\vert p^{\mathrm{R}%
}\right\vert \rightarrow0$) or the range $\Omega_{4}$ ($\left\vert
p^{\mathrm{L}}\right\vert \rightarrow0$),%
\begin{equation}
N_{n}^{\mathrm{cr}}\sim\left\vert p^{\mathrm{R}}\right\vert \rightarrow
0,\ \ N_{n}^{\mathrm{cr}}\sim\left\vert p^{\mathrm{L}}\right\vert
\rightarrow0,\ \ \forall\pi_{\bot}\neq0. \label{Nb}%
\end{equation}
This is exactly the behavior we see for all field configurations which are known
as exact solvable models \cite{x-case,L-field,x-exp}.

Absolute values of the asymptotic momenta $\left\vert p^{\mathrm{L}%
}\right\vert $ and $\left\vert p^{\mathrm{R}}\right\vert \ $are determined by
the quantum numbers $p_{0}$ and $p_{\bot}$; see Eq. (\ref{L3}). This fact
imposes a certain relation between both quantities. In particular, one can see
that $d\left\vert p^{\mathrm{L}}\right\vert /d\left\vert p^{\mathrm{R}%
}\right\vert <0,$ and at any given $p_{\bot}$, these quantities are restricted
inside the range $\Omega_{3}$,%
\begin{equation}
0\leq\left\vert p^{\mathrm{R/L}}\right\vert \leq p^{\mathrm{\max}%
},\ \ p^{\mathrm{\max}}=\sqrt{\Delta U\left(  \Delta U-2\pi_{\bot}\right)  }.
\label{d8}%
\end{equation}
It implies that%
\begin{equation}
0\leq\left\vert \left\vert p^{\mathrm{L}}\right\vert -\left\vert
p^{\mathrm{R}}\right\vert \right\vert \leq p^{\mathrm{\max}}. \label{g8}%
\end{equation}
Then for all $p_{0}$ and $p_{\bot}$ of the range $\Omega_{3}$, the numbers
$N_{n}^{\mathrm{cr}}$ are small and tend to zero if the Klein zone shrinks to
zero,%
\begin{equation}
N_{n}^{\mathrm{cr}}\sim\left\vert p^{\mathrm{R}}p^{\mathrm{L}}\right\vert
\rightarrow0\ \ \mathrm{if}\ \ p^{\mathrm{\max}}\rightarrow0. \label{tiny}%
\end{equation}
In this case, the probability of a pair creation with quantum numbers $n$,
$P(+-|0)_{n,n}$, can be approximated by the mean number $N_{n}^{\mathrm{cr}}$
as%
\begin{equation}
P(+-|0)_{n,n}=\frac{N_{n}^{\mathrm{cr}}}{1-N_{n}^{\mathrm{cr}}}P_{\mathrm{v}%
}\approx N_{n}^{\mathrm{cr}}. \label{pr}%
\end{equation}

The total number of created particles, $N^{\mathrm{cr}}$, given by the sum
(\ref{Ncr}) over such a tiny Klein zone, is small too. It tends to zero if the
magnitude of the potential step tends to a critical point,%
\begin{equation}
N^{\mathrm{cr}}\rightarrow0\ \ \mathrm{if}\ \ \Delta U\rightarrow2m.
\label{cr_p}%
\end{equation}

We see that $N^{\mathrm{cr}}\ll1$ near the critical point, so the probability of a
vacuum to remain a vacuum $P_{\mathrm{v}}$, given by a general form (\ref{Pv}),
can be approximated by the total number of created particles as $P_{\mathrm{v}%
}\approx1-N^{\mathrm{cr}}$. On the other hand, this probability can be
represented via the imaginary part of a one-loop effective action $S$ by the
seminal Schwinger formula,%
\begin{equation}
P_{\mathrm{v}}=\exp\left(  -2\mathrm{Im}S\right)  . \label{np1a}%
\end{equation}
Taking into account that $P_{\mathrm{v}}\approx1-2\mathrm{Im}S$, one finds a
relation%
\begin{equation}
2\mathrm{Im}S\approx N^{\mathrm{cr}}\ \ \mathrm{if\ \ }N^{\mathrm{cr}}%
\ll1\text{.} \label{np1b}%
\end{equation}
Taking into account this relation we can confirm the behavior (\ref{cr_p}) by the
results \cite{cr-regime1,cr-regime2} recently obtained for $\mathrm{Im}S$ in
inhomogeneous $x$-potential electric steps of an arbitrary configuration that
decay asymptotically with a power law $\sim E_{0}\left(  kx\right)  ^{-a}$,
$a\geq2$ or vanish at a finite point $\sim E_{0}\left(  k\left\vert
x-x_{0}\right\vert \right)  ^{b}$, $b>0$ (by taking the limit $a\rightarrow
\infty$ or $b\rightarrow\infty$, an exponentially decaying field is recovered),
where $E_{0}$ is a characteristic field strength scale and $k$ is a
characteristic length scale of the inhomogeneous field. It was shown by using
semiclassical worldline instanton methods \cite{instantons} in the weak-field
( $m^{2}/eE_{0}>1$) critical regime \cite{cr-regime1} and by analysis of
solutions of the Klein-Gordon and Dirac equation in the immediate vicinity of
the critical point for an arbitrary peak field strength \cite{cr-regime2} that
near criticality pair production vanishes, exhibiting universal properties
similar to those of continuous phase transitions.

In what follows, we consider a completely different type of universality,
believing that the Klein zone is quite extensive, so that the total number of
created pairs itself can be considered a large parameter. We face such
situations in astrophysical and condensed matter problems where the electric
field is strong. In these scenarios the numbers of the pairs created can reach
their limiting values, $N_{n}^{\mathrm{cr}}\rightarrow1$, and the total number
of pairs created, $N^{\mathrm{cr}}$, is not a small value anymore. For the
weakly inhomogeneous fields, this number is proportional to the large parameter
$L_{\mathrm{eff}}/\Delta l_{\mathrm{st}}$. For an arbitrary weakly inhomogeneous
strong electric field, one can derive in the leading-term approximation a
universal form for the total density of created pairs.

As it was explained above, the contributions to the total number of created
particles due to the part of the Klein zone in the vicinity of the critical
point are small and can be neglected in a following approximation. The main
contribution to $N^{\mathrm{cr}}$ is formed in some inner subrange
$D(x)$ of the Klein zone where transversal momentum $\pi_{\perp}$ and energy
$p_{0}$ are small enough. This inner subrange $D(x)$ can be described as%
\begin{equation}
D(x):\ \left\vert \pi_{0}(x)\right\vert \gg\pi_{\perp},\ \ \pi_{0}%
(x)=p_{0}-U(x). \label{np.2}%
\end{equation}
In $D(x)$, the effective particle energy is primarily determined by an
increment of energy $U(x)-U_{\mathrm{L}}$ or $U_{\mathrm{R}}-U(x)$ on the
spatial intervals $\Delta x=x-x_{\mathrm{L}}$ or $\Delta x=x_{\mathrm{R}}-x$,
respectively. It should be noted that $D(x)\subset D(x^{\prime})$ if
$x^{\prime}>x$.

Suppose that the electric field does not grow and decay abruptly at the edges
of some final interval, that is, the field slowly weakens at $x\rightarrow
\pm\infty$, and one of the points $x_{\mathrm{L}}$ or $x_{\mathrm{R}}$ or both
are infinitely distant from the origin, $x_{\mathrm{L}}\rightarrow-\infty$ and
$x_{\mathrm{R}}\rightarrow\infty$. In this case, the contributions to the flux
density $\tilde{n}^{\mathrm{cr}}$, given by Eq. (\ref{sc.21}), from the
regions $\left(  x_{\mathrm{L}},x_{\mathrm{eff}}^{\text{\textrm{in}}}\right]
$ and $\left(  x_{\mathrm{eff}}^{\text{\textrm{out}}},x_{\mathrm{R}}\right)  $
are exponentially small and can be disregarded, since the electric field in
these regions is very weak in comparison with the maximum value of the peak
field $E_{0}$, $E(x_{\mathrm{eff}}^{\text{\textrm{in}}}),E(x_{\mathrm{eff}%
}^{\text{\textrm{out}}})\ll E_{0}$. Therefore, in general, it is sufficient to
consider only the finite interval $\left(  x_{\mathrm{eff}}^{\text{\textrm{in}%
}},x_{\mathrm{eff}}^{\text{\textrm{out}}}\right]  $. We can divide this
interval into $M$ intervals,
\begin{align}
&  \ \Delta l_{i}=x_{i+1}-x_{i},\ \ i=1,\ldots,M,\ \nonumber\\
&  \ \sum_{i=1}^{M}\Delta l_{i}=x_{\mathrm{eff}}^{\text{\textrm{out}}%
}-x_{\mathrm{eff}}^{\text{\textrm{in}}},\ \ x_{1}=x_{\mathrm{eff}%
}^{\text{\textrm{in}}},\ x_{M+1}=x_{\mathrm{eff}}^{\text{\textrm{out}}%
}\text{,} \label{np.3}%
\end{align}
in such a way that Eqs.~(\ref{sc.1}) and (\ref{sc.2}) hold true for each of
these intervals. Let us show that this allows us to treat the electric field as approximately
uniform in each interval $\Delta l_{i}$, $\overline{E(x)}\approx \overline{E}%
(x_{i})$\ for $x\in \left( x_{i},x_{i+1}\right] $\ despite the fact that at
the beginning of each interval $\Delta l_{i}$\ the electric field $E(x)$\
changes abruptly. Note that it is possible to use, for example, sharp
exponential steps for the regularization of rectangular steps (see the details in
Ref. \cite{x-exp}) if the length of the interval where this change occurs is
significantly smaller than the length of each corresponding interval $\Delta
l_{i}$. However, as we can see, it is not necessary. 

In the case of the strong $L$-constant field, $m^{2}\lesssim eE_{0}$, and
the large parameter $\sqrt{eE_{0}}L$, a rough estimation of the
next-to-leading-term for the flux density of created pairs shows (see
details in Ref. \cite{L-field}) that it produces a small factor of
the order of $\left( \sqrt{eE_{0}}L\right) ^{-1}$, i.e.,
\begin{equation}
n^{\mathrm{cr}}=\tilde{n}^{\mathrm{cr}}\left[ 1+\frac{O(K)}{\sqrt{eE_{0}}L}%
\right] ,  \label{next}
\end{equation}%
where $\tilde{n}^{\mathrm{cr}}$\ is given by expression (ii) in Eq. (\ref{sc.23}). It
is clear that the abrupt change of the $L$-constant field at $x_{\mathrm{L}%
}=-L/2$ and $x_{\mathrm{R}}=L/2$\ entails considerable oscillations in the
distributions. Comparing the case of the $L$-constant field with other
examples of the exactly solvable cases \cite{x-case,x-exp}, we see that it
presents the roughest estimate of the neglected contributions for weakly
inhomogeneous potential steps. In particular, considering dominant
contributions to the flux density of pairs created by a very asymmetric
exponential peak (a field that grows from zero to its maximum value very rapidly and then experiences a smooth decay with the large effective length $L_{\mathrm{%
eff}}$) one can see that it does not depend on the details of the field
growth for the case of a strong field \cite{x-exp}. Then we can conclude that
this abrupt change cannot significantly influence the total value of $\tilde{%
n}^{\mathrm{cr}}$ as $N_{n}^{\mathrm{cr}}\leq 1$ for fermions. 

In each interval $\Delta l_{i}$, $\overline{E(x)}\approx \overline{E}(x_{i})$%
\ for $x\in \left( x_{i},x_{i+1}\right] $\ assuming that $L=\Delta l_{i}$,
we can approximate partial contribution to the flux density due to this
interval, $\Delta n_{i}^{\mathrm{cr}}$, as $\Delta n_{i}^{\mathrm{cr}%
}=\Delta \tilde{n}_{i}^{\mathrm{cr}}+O(K_{i})$, where $K_{i}$\ is any given
number satisfying the condition%
\begin{equation*}
\sqrt{e\overline{E}(x_{i})}\Delta l_{i}\gg K_{i}\gg \max \left\{ 1,m^{2}/e%
\overline{E}(x_{i})\right\} .
\end{equation*}%
Then, using Eqs. (\ref{sc.22}) and (\ref{sc.23}) for the $L$-constant
field,\ we have for $\tilde{n}^{\mathrm{cr}}$\ that%
\begin{align}
& \tilde{n}^{\mathrm{cr}}=\sum_{i=1}^{M}\Delta \tilde{n}_{i}^{\mathrm{cr}},\
\ \Delta \tilde{n}_{i}^{\mathrm{cr}}\approx \frac{J_{(d)}}{(2\pi )^{d-1}}%
\int_{ex_{i}\overline{E}(x_{i})}^{e\left( x_{i}+\Delta
l_{i}\right)\overline{E}(x_{i}) }dp_{0}\int_{\sqrt{\lambda _{i}}<K_{\bot }^{(i)}}d\mathbf{p}%
_{\bot }N_{n}^{(i)},  \notag \\
& N_{n}^{(i)}=e^{-\pi \lambda _{i}},\ \ \lambda _{i}=\pi _{\bot }^{2}/e%
\overline{E}(x_{i}),  \label{np.4}
\end{align}%
where $K_{\bot }^{(i)}$ are any given numbers satisfying the conditions%
\emph{\ }%
\begin{equation*}
 K_{i}\gg \left[ K_{\bot }^{(i)}%
\right] ^{2}\gg \max \left\{ 1,m^{2}/e\overline{E}(x_{i})\right\} .
\end{equation*}
 We can formally represent the variable $p_{0}$ in the latter expression as%
\begin{equation}
p_{0}=U(x),\ U(x)=\int_{x_{\mathrm{L}}}^{x}dx^{\prime}\ eE(x^{\prime
})+U_{\mathrm{L}},\ \ dp_{0}=eE(x)dx. \label{np.5}%
\end{equation}
Then neglecting small contributions to the integral (\ref{np.4}), we find the
following universal form for the flux density of created pairs in the
leading-term approximation for a weakly inhomogeneous, but otherwise arbitrary
strong electric field%
\begin{equation}
\tilde{n}^{\mathrm{cr}}\approx\frac{J_{(d)}}{(2\pi)^{d-1}}\int_{x_{\mathrm{L}%
}}^{x_{\mathrm{R}}}dx\ eE(x)\int d\mathbf{p}_{\bot}N_{n}^{\mathrm{uni}%
},\ \ N_{n}^{\mathrm{uni}}=\exp\left[  -\pi\frac{\pi_{\bot}^{2}}%
{eE(x)}\right]  . \label{np.6}%
\end{equation}
The quantity $N_{n}^{\mathrm{uni}}$ has a universal form which can be used to
calculate any total characteristic of the pair creation effect. One can
integrate the latter expression over $d\mathbf{p}_{\bot}$ to obtain the final
form,
\begin{equation}
\tilde{n}^{\mathrm{cr}}\approx\frac{J_{(d)}}{(2\pi)^{d-1}}\int_{x_{\mathrm{L}%
}}^{x_{\mathrm{R}}}dx\ \left[  eE(x)\right]  ^{d/2}\exp\left[  -\pi\frac
{m^{2}}{eE(x)}\right]  . \label{np.7}%
\end{equation}
These universal forms can be derived for bosons as well, if we are restricting
them to the forms of external electric fields, namely, fields that have no abrupt
variations of $E(x)$ that can produce significant growth of $N_{n}%
^{\mathrm{cr}}$ on a finite spatial interval; i.e., we have to include in the
range $D$, the only subranges where $N_{n}^{\mathrm{cr}}\leq1$. In this case,
the universal forms for bosons are the same, Eqs. (\ref{np.6}) and
(\ref{np.7}), with $J_{(d)}=1$ for scalar particles and $J_{(d)}=3$ for vector ones.

Using the identity $-\ln\left(  1-N_{n}^{\mathrm{cr}}\right)  =N_{n}%
^{\mathrm{cr}}+\left(  N_{n}^{\mathrm{cr}}\right)  ^{2}+\ldots$, in the same
manner we can derive a universal form of the probability of a vacuum to remain a
vacuum $P_{\mathrm{v}}$ defined for fermions by Eq. (\ref{Pv}). First, we get
\begin{equation}
P_{\mathrm{v}}\approx\exp\left\{  -\frac{V_{\bot}TJ_{(d)}}{(2\pi)^{d-1}}%
\sum_{l=1}^{\infty}\int_{x_{\mathrm{L}}}^{x_{\mathrm{R}}}dxeE(x)\int
d\mathbf{p}_{\bot}\left(  N_{n}^{\mathrm{uni}}\right)  ^{l}\right\}  .
\label{np.8}%
\end{equation}
After integration over $\mathbf{p}_{\bot}$, we finally obtain%
\begin{equation}
P_{\mathrm{v}}\approx\exp\left\{  -\frac{V_{\bot}TJ_{(d)}}{(2\pi)^{d-1}}%
\sum_{l=1}^{\infty}\int_{x_{\mathrm{L}}}^{x_{\mathrm{R}}}dx\frac{\left[
eE(x)\right]  ^{d/2}}{l^{d/2}}\exp\left[  -\pi\frac{lm^{2}}{eE(x)}\right]
\right\}  . \label{np.9}%
\end{equation}

For bosons, we know that the vacuum-to-vacuum transition probability has the form,
\begin{equation}
P_{\mathrm{v}}^{(\mathrm{boson})}=\exp\left[  -\sum_{n}\ln\left(
1+N_{n}^{\mathrm{cr}}\right)  \right]  , \label{np.10}%
\end{equation}
so the universal form of the vacuum-to-vacuum transition probability for the
Bose case is%
\begin{equation}
P_{\mathrm{v}}^{(\mathrm{boson})}\approx\exp\left\{  -\frac{V_{\bot}TJ_{(d)}%
}{(2\pi)^{d-1}}\sum_{l=1}^{\infty}\int_{x_{\mathrm{L}}}^{x_{\mathrm{R}}%
}dx\left(  -1\right)  ^{l-1}\frac{\left[  eE(x)\right]  ^{d/2}}{l^{d/2}}%
\exp\left[  -\pi\frac{lm^{2}}{eE(x)}\right]  \right\}  , \label{np.11}%
\end{equation}
where $J_{(d)}$ is the number of boson spin degrees of freedom.

Using Eqs. (\ref{np.7}) and (\ref{np.9}), one can precisely reproduce
expressions (\ref{sc.23}) and (\ref{sc.27}) that are found for the total
densities and the vacuum-to-vacuum transition probabilities when directly
adopting the weakly inhomogeneous field approximation to the exactly solvable
cases. Comparing Eqs. (\ref{np.7}) and (\ref{np.11}) with the exact results
obtained for bosons \cite{x-case,L-field,x-exp}, one finds precise agreement
too. Thus, we have a confirmation of the universal forms obtained above.

The representations (\ref{np.9}) and (\ref{np.11}) coincide with the
vacuum-to-vacuum transition probabilities obtained from the imaginary part of
a locally constant field approximation (LCFA) for the one-loop effective
action in $d=4$ dimensions \cite{GiesK17,Karb17}. In this approximation, the
effective action $S$ is expanded about the constant field case, in terms of
derivatives of the background field strength $F_{\mu\nu}$,%
\begin{equation}
S=S^{\left(  0\right)  }[F_{\mu\nu}]+S^{\left(  2\right)  }[F_{\mu\nu
},\partial_{\mu}F_{\nu\rho}]+... \label{uni7b}%
\end{equation}
where $S^{\left(  0\right)  }$ involves no derivatives of the background field
strength $F_{\mu\nu}$ (that is, $S^{\left(  0\right)  }$ is a locally constant
field approximation for $S$ that has a form of the Heisenberg-Euler action),
while the first correction $S^{\left(  2\right)  }$ involves two derivatives
of the field strength, and so on; see Ref.~\cite{Dunn04} for a review. Using
the representation (\ref{np1a}), one finds the LCFA for the probability
$P_{\mathrm{v}}$ as
\begin{equation}
P_{\mathrm{v}}\approx\exp\left(  -2\mathrm{Im}S^{\left(  0\right)  }\right)  .
\label{LCFA}%
\end{equation}
However, it should be stressed that unlike the representations obtained in
Refs.~\cite{GiesK17,Karb17}, we derive Eqs.~(\ref{np.9}) and (\ref{np.11}) in
the framework of the general formulation of strong-field QED in the presence
of $x$-electric potential steps \cite{x-case}, where $P_{\mathrm{v}}$ are
defined by Eqs.~(\ref{Pv}) and (\ref{np.10}), respectively. Therefore, we
obtain Eqs.~(\ref{np.9}) and (\ref{np.11}) independently from the derivative
expansion approach, and the obtained result holds true for any strong field
under consideration. It is known that for a general background field, it is
extremely difficult to estimate and compare the magnitude of various terms in
the derivative expansion. Only under the assumption $m^{2}/eE_{0}>1$, one can
demonstrate that the derivative expansion is completely consistent with the
semiclassical WKB analysis of the imaginary part of the effective action
\cite{DunnH98}. Thus, the representations (\ref{np.9}) and (\ref{np.11}) are
proof that the imaginary part of the LCFA for the Heisenberg-Euler action is
correct for an arbitrarily weakly inhomogeneous electric field of a constant
direction. The universal forms (\ref{np.6}) and (\ref{np.7}) for the flux
density of created pairs are completely new and present a LCFA for this
physical quantity. It is a new kind of a LCFA obtained without any relation to
the Heisenberg-Euler action.

One can see that the obtained universal forms have especially simple forms in
two limited cases, for a weak electric field ($m^{2}/eE_{0}\gg1$), when the
term $\left[  eE(x)\right]  ^{d/2}$ can be approximated by its maximum value,
$\left[  eE_{0}\right]  ^{d/2}$, and a strong electric field ($m^{2}/eE_{0}\ll
1$), when there exist spatial intervals where $m^{2}/eE(x)\ll1$ and
approximations of the type%
\begin{equation}
\exp\left[  -\frac{\pi lm^{2}}{eE(x)}\right]  =1-\frac{\pi lm^{2}}%
{eE(x)}+\ldots\label{np.12}%
\end{equation}
are available. For example, one can consider the case of a strong Gauss peak,%
\begin{equation}
E(x)=E_{0}\exp\left[  -\left(  x/L_{\mathrm{G}}\right)  ^{2}\right]  ,
\label{np.13}%
\end{equation}
with a large parameter $L_{\mathrm{G}}\rightarrow\infty$. In this case, we do
not have an exact solution of the Dirac equation, and known semiclassical
approximations (valid for a weak field) are not applicable. However, using
approximation (\ref{np.12}), we find from Eqs. (\ref{np.7}) and (\ref{np.9}),
the leading term as%
\begin{equation}
\tilde{n}^{\mathrm{cr}}\approx\frac{J_{(d)}\left(  eE_{0}\right)
^{d/2}L_{\mathrm{G}}}{d(2\pi)^{d-2}},\ \ P_{\mathrm{v}}\approx\exp\left(
-V_{\bot}T\tilde{n}^{\mathrm{cr}}\sum_{l=1}^{\infty}l^{-d/2}\right)  .
\label{np.14}%
\end{equation}

As another example, we consider an inverse square electric field,
\begin{equation}
E(x)=E_{0}\left[  1+\left(  \frac{2x}{L_{\mathrm{w}}}\right)  ^{2}\right]
^{-1},\ \ A_{0}(x)=-\frac{L_{\mathrm{w}}}{2}E_{0}\ \mathrm{arctg}\frac
{2x}{L_{\mathrm{w}}}. \label{np.15}%
\end{equation}
It is a particular case of an inhomogeneous field that was used to study the
LCFA for the probability $P_{\mathrm{v}}$ in Ref.~\cite{Karb17}. Using the
Eqs. (\ref{np.7}) and (\ref{np.9}), we find in the leading order that%
\begin{align}
&  \tilde{n}^{\mathrm{cr}}\approx\frac{L_{\mathrm{w}}}{2}r^{\mathrm{cr}}%
\delta_{\mathrm{w}},\label{np.16a}\\
&  P_{\mathrm{v}}\approx\exp\left\{  -V_{\bot}T\tilde{n}^{\mathrm{cr}}%
\sum_{l=0}^{\infty}\frac{\epsilon_{l+1}}{\left(  l+1\right)  ^{d/2}}%
\exp\left[  -\frac{\pi lm^{2}}{eE_{0}}\right]  \right\}  , \label{np.16b}%
\end{align}
where
\begin{equation}
\epsilon_{l}=\epsilon_{l}^{\mathrm{w}}=\sqrt{\pi}\Psi(\frac{1}{2},\frac
{3-d}{2};\frac{\pi lm^{2}}{eE_{0}})\delta_{\mathrm{w}}^{-1},\ \ \delta
_{\mathrm{w}}=\sqrt{\pi}\Psi(\frac{1}{2},\frac{3-d}{2};\frac{\pi m^{2}}%
{eE_{0}}). \label{np.17}%
\end{equation}
Our result for the probability $P_{\mathrm{v}}$ (\ref{np.16b}) coincides with the
one obtained in Ref.~\cite{Karb17} for the particular case of $d=4$.

For the case when the external background is given by a slowly varying in time
electric field, the expressions for the total number of created particles and
for the probability of vacuum-to-vacuum transition similar to Eqs.
(\ref{np.7}) and (\ref{np.9}) were obtained in Ref. \cite{GavGit17}. It is
clear, however, that time-dependent fields and nonuniform fields describe
physically distinct situations, and in the general case the results have different forms.

\section{Mean current \label{S4}}

It is well known that in QFT, measurable values and their mean values,
generally speaking, are defined globally. This fact implies that those values
are defined in some macroscopic volume at some fixed moment of time. Of
course, the measurement procedure takes some macroscopic time itself. This is
usually ignored under the assumption that the measured physical value does not
change substantially during the measurement. If an external field does not
violate vacuum stability, then it just causes vacuum polarization inside the
area occupied by the field. This effect is a quasilocal one. In the case that
is most interesting to us, i.e., when an external field is capable of violating
vacuum stability, there is a global effect of the field due to an
electron-positron pair creation. These pairs do not disappear when the field
is turned off. The electrons and positrons created leave the area
$S_{\mathrm{int}}$ occupied by the field, creating a constant flow of
particles moving away from the field. For symmetry reasons, it is clear that
this flow is aligned along the field direction (which is the axis $x$ in our
case) and creates a longitudinal electric current.

Created electrons and positrons leaving the field region $S_{\mathrm{int}}$
fly out into the regions $S_{\mathrm{L}}$ and $S_{\mathrm{R}}$,
respectively, and move away from the electric field at constant
longitudinal velocities $-v^{\mathrm{L}}$ and $v^{\mathrm{R}}$,
where $v^{\mathrm{L}}=\left\vert p^{\mathrm{L}}/\pi_{0}\left(  \mathrm{L}%
\right)  \right\vert $ and $v^{\mathrm{R}}=\left\vert p^{\mathrm{R}}/\pi
_{0}\left(  \mathrm{R}\right)  \right\vert $. This longitudinal current is
coordinate independent and, therefore, can be determined by its value anywhere
in $S_{\mathrm{L}}$ or $S_{\mathrm{R}}$.

Using a general theory \cite{x-case}, we can derive the forms for the vacuum
mean current. The charge operator $\hat{Q}$ is defined in Ref. \cite{x-case}
as a commutator,
\begin{equation}
\hat{Q}=-\frac{e}{2}\int \left[ \hat{\Psi}^{\dagger }\left( X\right) ,\hat{%
\Psi}\left( X\right) \right] _{-}d\mathbf{r\ }.  \label{Charge}
\end{equation}%
The expression (\ref{Charge}) for the charge operator suggests the
definition for the current operator,
\begin{equation}
\hat{J}^{\mu }=-\frac{e}{2}\left[ \hat{\Psi}^{\dag }(X)\gamma ^{0}\gamma ^{\mu },\ 
\hat{\Psi}(X)\right]_{-} \label{Current}.
\end{equation}
These values can be expressed via the vacuum mean values of the
electric current density of the Dirac field through the surface
$x=\mathrm{const}$, which are defined as follows,%
\begin{align}
&  \left\langle J^{\mu}(x)\right\rangle _{\mathrm{in}}=\left\langle
0,\mathrm{in}\right\vert \hat{J}^{\mu}\left\vert 0,\mathrm{in}\right\rangle
,\ \ \left\langle J^{\mu}(x)\right\rangle _{\mathrm{out}}=\left\langle
0,\mathrm{out}\right\vert \hat{J}^{\mu}\left\vert 0,\mathrm{out}\right\rangle
,\label{m1}
\end{align}
where the Dirac Heisenberg operator $\hat{\Psi}\left(  X\right)  $ is assigned
to the Dirac field $\psi\left(  X\right)  $. We stress the $x$ coordinate
dependence of mean values (\ref{m1}), which does exist due to the coordinate
dependence of the external field. It should be noted that these densities
depend on a vacuum definition and the structure of the electric field in the
direction\emph{\ }$x$. The renormalized vacuum mean value $\left\langle
J^{\mu}(x)\right\rangle _{\mathrm{in}}$ is a source in the equations of motion for a
mean electromagnetic field. The quantity $\left\langle J^{\mu}(x)\right\rangle
_{\mathrm{out}}$ is needed to present the operator $\hat{J}^{\mu}$ in a normal ordering
form with respect to the creation and annihilation operators of the final
particles,%
\begin{equation}
\hat{N}_{\mathrm{out}}\left(  J^{\mu}\right)  =\hat{J}^{\mu}-\left\langle J^{\mu
}(x)\right\rangle _{\mathrm{out}}. \label{m2}%
\end{equation}

Mean values and probability amplitudes are described by the Feynman diagrams
with two kinds of charged particle propagators in the external field under
consideration, respectively. The probability amplitudes are calculated using
the causal (Feynman) propagator $S^{c}(X,X^{\prime})$ while mean values are
found using the so-called in-in propagator $S_{\text{\textrm{in}}}%
^{c}(X,X^{\prime})$ and out-out propagator $S_{\text{\textrm{out}}}%
^{c}(X,X^{\prime})$,
\begin{align}
&  S^{c}(X,X^{\prime})=i\left\langle 0,\mathrm{out}\right\vert \hat{T}%
\hat{\Psi}(X)\hat{\Psi}^{\dag}(X^{\prime})\gamma^{0}\left\vert 0,\mathrm{in}%
\right\rangle c_{\mathrm{v}}^{-1},\nonumber\\
&  S_{\text{\textrm{in}}}^{c}(X,X^{\prime})=i\left\langle 0,\mathrm{in}%
\right\vert \hat{T}\hat{\Psi}(X)\hat{\Psi}^{\dag}(X^{\prime})\gamma
^{0}\left\vert 0,\mathrm{in}\right\rangle ,\nonumber\\
&  S_{\text{\textrm{out}}}^{c}(X,X^{\prime})=i\left\langle 0,\mathrm{out}%
\right\vert \hat{T}\hat{\Psi}(X)\hat{\Psi}^{\dag}(X^{\prime})\gamma
^{0}\left\vert 0,\mathrm{out}\right\rangle , \label{m5.1}%
\end{align}
where $\hat{T}$ denotes the chronological ordering operation and
$c_{\mathrm{v}}$ is the vacuum-to-vacuum transition amplitude,{\large \ }%
$c_{\mathrm{v}}=\left\langle 0,\mathrm{out}\right\vert \left.  0,\mathrm{in}%
\right\rangle ${\large ,\ }$\left\vert c_{\mathrm{v}}\right\vert
^{2}=P_{\mathrm{v}}${\large .} As usual, these propagators can be expressed
via the following singular functions%
\begin{align}
&  S^{c}(X,X^{\prime})=\theta(t-t^{\prime})\,S^{-}\left(  x,x^{\prime}\right)
-\theta(t^{\prime}-t)\,S^{+}\left(  x,x^{\prime}\right)  ,\nonumber\\
&  S_{\mathrm{in/out}}^{c}(X,X^{\prime})=\theta(t-t^{\prime}%
)S_{\mathrm{in/out}}^{-}(X,X^{\prime})-\theta(t^{\prime}-t)S_{\mathrm{in/out}%
}^{+}(X,X^{\prime})\,. \label{m5.2}%
\end{align}
The vacuum mean values (\ref{m1}) can be expressed via the propagators
$S_{\text{\textrm{in}}}^{c}$ and $S_{\text{\textrm{out}}}^{c}$ while the
causal propagator $S^{c}$ determines the vacuum polarization contribution to
current as%
\begin{align}
&  \left\langle J^{\mu}(x)\right\rangle ^{c}=\left\langle 0,\mathrm{out}%
\right\vert \hat{J}^{\mu}\left\vert 0,\mathrm{in}\right\rangle c_{\mathrm{v}}%
^{-1}=-ie\mathrm{tr}\left[  \gamma^{\mu}S^{c}(X,X^{\prime})\right]
|_{X=X^{\prime}},\nonumber\\
&  \left\langle J^{\mu}(x)\right\rangle _{\mathrm{in/out}}=-ie\mathrm{tr}%
\left[  \gamma^{\mu}S_{\text{\textrm{in/out}}}^{c}(X,X^{\prime})\right]
|_{X=X^{\prime}}. \label{m5.3}%
\end{align}

Using the explicit forms of these singular functions, given by in Ref.
\cite{x-case}, we see that transversal components of these currents are equal
to zero,%
\begin{equation}
\left\langle J^{k}(x)\right\rangle _{\mathrm{in}}=\left\langle J^{k}%
(x)\right\rangle _{\mathrm{out}}=\left\langle J^{k}(x)\right\rangle
^{c}=0\ \ \mathrm{if}\ \ k\neq1,\ \label{m6}%
\end{equation}
due to the cylindrical symmetry of the problem.

Using a singular function,
\begin{equation}
S^{p}(X,X^{\prime})=S_{\text{\textrm{in}}}^{c}(X,X^{\prime})-S^{c}%
(X,X^{\prime}) \label{Sp}%
\end{equation}
it is useful to introduce a current,
\begin{equation}
\left\langle J^{\mu}(x)\right\rangle ^{p}=-ie\mathrm{tr}\left[  \gamma^{\mu
}S^{p}(X,X^{\prime})\right]  |_{X=X^{\prime}}\ . \label{m13.1}%
\end{equation}
The explicit form of $S^{p}(X,X^{\prime})$ is \cite{x-case}
\begin{align}
&  S^{p}(X,X^{\prime})=i\sum_{n\in\Omega_{3}}\mathcal{M}_{n}^{-1}\left[
g\left(  _{-}\left\vert ^{-}\right.  \right)  ^{\ast}\right]  ^{-1}\ _{-}%
\psi_{n}\left(  X\right)  \ ^{-}\bar{\psi}_{n}\left(  X^{\prime}\right)
,\nonumber\\
&  \mathcal{M}_{n}=2\frac{\tau^{\left(  \mathrm{R}\right)  }}{T}\left\vert
g\left(  _{+}\left\vert ^{-}\right.  \right)  \right\vert ^{2}=2\frac
{\tau^{\left(  \mathrm{L}\right)  }}{T}\left\vert g\left(  _{+}\left\vert
^{-}\right.  \right)  \right\vert ^{2}. \label{m13.2}%
\end{align}
Note that $S^{p}$ is formed in the range $\Omega_{3}$ only and vanishes if
there is no pair creation. Here, $\tau^{\left(  \mathrm{L}\right)  }$ and
$\tau^{\left(  \mathrm{R}\right)  }$ are equal macroscopic times of motion for
created particles in the regions $S_{\mathrm{L}}$ and $S_{\mathrm{R}}$,
respectively. It is assumed that the regions $S_{\mathrm{L}}$ and
$S_{\mathrm{R}}$ are substantially wider than the region $S_{\mathrm{int}}$,
and it is possible to neglect the time period when the created particles are
moving from the region $S_{\mathrm{int}}$ into regions $S_{\mathrm{L}}$ and
$S_{\mathrm{R}}$. We suppose that all the measurements are performed during a
macroscopic time $T$ when the external field can be considered as constant. In
particular, a charge transport through planes $x=x_{\mathrm{L}}$ and
$x=x_{\mathrm{R}}$ occurs during the time period $T$. In this case, the times
$\tau^{\left(  \mathrm{L}\right)  }$ and $\tau^{\left(  \mathrm{R}\right)  }$
coincide with $T$, $\tau^{\left(  \mathrm{L}\right)  }=\tau^{\left(
\mathrm{R}\right)  }=T$, and we obtain that%
\[
\mathcal{M}_{n}^{-1}=\frac{1}{2}N_{n}^{\mathrm{cr}},
\]
where $N_{n}^{\mathrm{cr}}$ is given by Eq. (\ref{meanN}).

Thus, we have that%
\begin{equation}
\left\langle J^{\mu}(x)\right\rangle _{\mathrm{in}}=\left\langle J^{\mu
}(x)\right\rangle ^{c}+\left\langle J^{\mu}(x)\right\rangle ^{p}.
\label{m13.3}%
\end{equation}
It is clear from Eq.~(\ref{m6}) that only the components $\left\langle
J^{0}(x)\right\rangle ^{p}$ and $\left\langle J^{1}(x)\right\rangle ^{p}$ are
nonzero. Using representations (\ref{m13.1}) and (\ref{m13.2}) and
decomposition (\ref{dec}), we find that%
\begin{align}
&  \left\langle J^{0}(x)\right\rangle ^{p}=\left\{
\begin{array}
[c]{l}%
-\bar{J}^{0}\left(  \mathrm{L}\right)  \ \ \mathrm{if}\ \ x\in S_{\mathrm{L}%
}\\
\bar{J}^{0}\left(  \mathrm{R}\right)  \ \ \mathrm{if}\ \ x\in S_{\mathrm{R}}%
\end{array}
\right.  ,\ \ \nonumber\\
&  \bar{J}^{0}\left(  \mathrm{L/R}\right)  =\frac{e}{2}\sum_{n\in\Omega_{3}%
}j_{n}^{0}\left(  \mathrm{L}/\mathrm{R}\right)  ,\ \ j_{n}^{0}\left(
\mathrm{L}/\mathrm{R}\right)  =j_{n}^{1}/v^{\mathrm{L/R}};\label{m17a}\\
&  \left\langle J^{1}(x)\right\rangle ^{p}=\frac{e}{2}\sum_{n\in\Omega_{3}%
}j_{n}^{1},\ \ j_{n}^{1}=N_{n}^{\mathrm{cr}}(TV_{\perp})^{-1}. \label{m17b}%
\end{align}

Using Eqs. (\ref{L3}), we can present the singular functions
$S_{\text{\textrm{in}}}^{c}$ and $S_{\text{\textrm{out}}}^{c}$ given by in Ref.~\cite{x-case}, in an
explicit form in the regions $S_{\mathrm{L}}$ and $S_{\mathrm{R}}$,
respectively. We find that%
\begin{equation}
\left\langle J^{1}(x)\right\rangle _{\mathrm{in}}=-\left\langle J^{1}%
(x)\right\rangle _{\mathrm{out}}=\left\langle J^{1}(x)\right\rangle
^{p}\ \ \mathrm{if}\ \ x\in S_{\mathrm{L}}\ \ \mathrm{or}\ \ S_{\mathrm{R}}.
\label{m7}%
\end{equation}
Note that the values of electric current densities (\ref{m7}) [including each
of the components of $j_{n}^{1}$, given by Eq. (\ref{m17b})] are conserved
along the axis $x$. These expressions are defined for regions where an
electric field is absent and do not contain contributions independent of an
electric field. For example, if an electric field in the region
$S_{\mathrm{int}}$ turns off, $E\rightarrow0$, then the number of pairs
created by the field vanishes, $N_{n}^{\mathrm{cr}}\rightarrow
0$\emph{.} For this reason, the densities (\ref{m7}) are characteristics of
real particles and cannot change after an electric field is turned off.

Taking into account the relations (\ref{m2}) and (\ref{m7}), we find that the
longitudinal current of pairs created is%
\begin{equation}
J_{\mathrm{cr}}^{1}=\left\langle N_{\mathrm{out}}\left(  J^{1}\right)
\right\rangle _{\mathrm{in}}=2\left\langle J^{1}(x)\right\rangle ^{p}%
=e\sum_{n\in\Omega_{3}}j_{n}^{1}. \label{m10}%
\end{equation}
Here, $j_{n}^{1}$ is the flux density of particles created with a given $n$, and
\begin{equation}
\sum_{n\in\Omega_{3}}j_{n}^{1}=n^{\mathrm{cr}} \label{m11}%
\end{equation}
is the total flux density of created particles, given by Eq. (\ref{sc.20}).
This allows us to interpret the density $2\left\langle J^{0}(x)\right\rangle
^{p}$ as the charge density of the particles created,%
\begin{equation}
J_{\mathrm{cr}}^{0}(x)=2\left\langle J^{0}(x)\right\rangle ^{p}=\left\{
\begin{array}
[c]{l}%
-e\sum_{n\in\Omega_{3}}j_{n}^{0}\left(  \mathrm{L}\right)  \ \ \mathrm{if\ }%
\ x\in S_{\mathrm{L}}\\
e\sum_{n\in\Omega_{3}}j_{n}^{0}\left(  \mathrm{R}\right)  \ \ \mathrm{if}%
\ \ x\in S_{\mathrm{R}}%
\end{array}
\right.  .\ \label{m19}%
\end{equation}

This interpretation also works for partial components of density
(\ref{m19}). We see that created electrons with a given $n$ move with a velocity
$v^{\mathrm{L}}$ in a direction opposite to the direction of the axis $x$,
i.e., in the direction opposite to the direction of the current density
$ej_{n}^{1}$. During the time $T$, that these electrons transport through the plane
$x=x_{\mathrm{L}}$, the amount of charge per $V_{\bot}$ is equal to $ej_{n}^{1}T$.
Taking into account that this charge is distributed uniformly over the
cylindrical volume with the length $v^{\mathrm{L}}T$, we obtain that the charge
density of created electrons with a given $n$ is equal to $ej_{n}^{1}/\left(
-v^{\mathrm{L}}\right)  =-ej_{n}^{0}\left(  \mathrm{L}\right)  $, where
$j_{n}^{0}\left(  \mathrm{L}\right)  $ is given by Eq. (\ref{m17a}). Created
positrons with a given $n$ move at a velocity $v^{\mathrm{R}}$ along axis $x$;
i.e., the direction of their movement coincides with the direction of the
current density $ej_{n}^{1}$. During the time $T$ positrons transport the same
charge amount per $V_{\bot}$ as electrons, $ej_{n}^{1}T$, through the plane
$x=x_{\mathrm{R}}$ (in this case it is uniformly distributed over a
cylindrical volume with the length $v^{\mathrm{R}}T$). We find that the charge
density of created positrons with a given $n$ is $ej_{n}^{1}/v^{\mathrm{R}%
}=ej_{n}^{0}\left(  \mathrm{R}\right)  $, where $j_{n}^{0}\left(
\mathrm{R}\right)  $ is given by Eq. (\ref{m17a}). We see that every pair
$ej_{n}^{1}$ and $-ej_{n}^{0}\left(  \mathrm{L}\right)  $ in $S_{\mathrm{L}}$
and $ej_{n}^{1}$ and $ej_{n}^{0}\left(  \mathrm{R}\right)  $ in $S_{\mathrm{R}%
}$ , correspondingly, can be connected by a Lorentz boost and represents
(nonzero) components of the same Lorentz vector. The number densities in both
regions $x\in S_{\mathrm{L}}$ and $x\in S_{\mathrm{R}}$ are equal,
\[
\sum_{n\in\Omega_{3}}j_{n}^{0}\left(  \mathrm{L}\right)  =\sum_{n\in\Omega
_{3}}j_{n}^{0}\left(  \mathrm{R}\right)  .
\]
Thus, the charge densities of created electrons in $x\in S_{\mathrm{L}}$ and
positrons in $x\in S_{\mathrm{R}}$ have the same value, but opposite sign. We see
that the total charge of created particles is zero, and an electric field
produces a charge polarization along axis $x$, just as one would expect.

The main contribution to the longitudinal current and flux density of the pair
created is formed in the inner subrange $D(x)$ of the Klein zone, given by the
inequality (\ref{np.2}). In this subrange, $v^{\mathrm{L}}\simeq v^{\mathrm{R}%
}\simeq1$, and we have that $j_{n}^{0}\left(  \mathrm{L/R}\right)  \simeq
j_{n}^{1}$. Then, we can write that%
\begin{equation}
J_{\mathrm{cr}}^{1}=e\tilde{n}^{\mathrm{cr}},\ \ J_{\mathrm{cr}}%
^{0}(x)=\left\{
\begin{array}
[c]{l}%
-e\tilde{n}^{\mathrm{cr}}\ \ \mathrm{if}\ \ x\in S_{\mathrm{L}}\\
e\tilde{n}^{\mathrm{cr}}\ \ \mathrm{if\ }\ x\in S_{\mathrm{R}}%
\end{array}
\right.  ,\ \label{m20}%
\end{equation}
where $\tilde{n}^{\mathrm{cr}}$ is given by the universal forms (\ref{np.6})
and (\ref{np.7}).

\section{Concluding remarks}

In the present article, we have presented the approximation that allows one to
treat nonper\-tur\-ba\-ti\-vely the vacuum instability effects for arbitrary weakly
inhomogeneous $x$-electric potential steps in the absence of the corresponding
exact solutions. First, we have revised vacuum instability effects in three
exactly solvable cases in QED with $x$-electric potential steps that have a
real physical importance. These are the Sauter electric field, the so-called
$L$-constant electric field, and the exponentially growing and decaying strong
electric weakly inhomogeneous fields. Defining the conditions of a field being
weakly inhomogeneous in general terms, we observed some universal features of
vacuum effects caused by the strong electric fields. These universal features
appear when the length of the external field is sufficiently large in
comparison to the scale, $\Delta l_{\mathrm{st}}=\left[  e\overline
{E(x)}\right]  ^{-1/2}$. In this case, the scale of the variation for an
external field and leading contributions to vacuum mean values are
macroscopic. We found universal approximate representations for the flux
density of created pairs (bosons and fermions) and the probability of the
vacuum to remain a vacuum in the leading-term approximation for a weakly
inhomogeneous, but otherwise arbitrary strong electric field. These
representations do not require a knowledge of corresponding solutions of the
Dirac equation; they have a form of simple functionals of a given weakly
inhomogeneous electric field. The universal forms for the flux density of
created pairs are completely new and present a LCFA for this physical
quantity. It is a new kind of a LCFA obtained without any relation to the
Heisenberg-Euler action. We established relations of these representations
with leading term approximations of derivative expansion results. We have
tested the obtained representations for cases of exactly solvable $x$-electric
potential steps (based on using exact solutions). We have also considered two
examples of $x$-electric potential steps where the exact solutions of the
corresponding Dirac equation are not known, a Gauss peak and an inverse square
electric field. We found the longitudinal current density and charge density
of created electrons and positrons and related these densities to the
corresponding flux density. In the regions $S_{\mathrm{L}}$ and $S_{\mathrm{R}%
}$, where the electric field is absent (or negligible), leading vacuum
characteristics are formed due to the real pair production. Thus, we have
isolated global contributions that depend on the total history of an electric
field from local contributions formed in the region $S_{\mathrm{int}}$. The nonperturbative (with respect to the external field) technique elaborated on in Ref.~\cite{x-case} allows one to calculate all the characteristics of zero-order processes and Feynman diagrams that describe all characteristics of processes with an interaction between charged particles and photons. These diagrams formally have the usual form but contain special propagators. Using expressions for these propagators in terms of in- and out-solutions, presented in Ref.~\cite{x-case}, our approximation method can be easily adapted to calculate one-loop and higher order contributions. The first step in developing the corresponding nonperturbative technique was recently done in Ref.~\cite{BrGavGitIv}, where a relation between the electron propagator in a constant electric field confined between two capacitor plates and the well-known Fock-Schwinger proper-time integral representation is established.

\section{Acknowledgement}

S.P.G. and D.M.G. acknowledge support from Tomsk State University
Competitiveness Improvement Program and the partial support from the Russian
Foundation for Basic Research (RFBR), under Project No. 18-02-00149;
D.M.G. is also supported by the Grant No. 2016/03319-6, Funda\c{c}\~{a}o de Amparo \`{a}
Pesquisa do Estado de S\~{a}o Paulo (FAPESP), and permanently by Conselho Nacional
de Desenvolvimento Cient\'{i}fico e Tecnol\'{o}gico (CNPq). The work of A.A.S. was
supported by Grant No. 2017/05734-3 of FAPESP.


\begin{thebibliography}{99}                                                                                               %

\bibitem {Schwinger51}J. Schwinger, Phys. Rev. \textbf{82}, 664 (1951).

\bibitem {Usov97}V. V. Usov, Phys. Rev. Lett. \textbf{80}, 230 (1998); D. Page
and V. V. Usov, Phys. Rev. Lett. \textbf{89}, 131101 (2002); V. V. Usov, Phys.
Rev. D \textbf{70}, 067301 (2004); J. Madsen, Phys. Rev.
Lett. \textbf{100}, 151102 (2008); A. Kurkela, P. Romatschke, and A. Vuorinen,
Phys. Rev. D \textbf{81}, 105021 (2010).

\bibitem {Alf+etal01}M. Alford, K. Rajagopal, S. Reddy, and F. Wilczek, Phys.
Rev. D \textbf{64}, 074017 (2001).

\bibitem {Ruf+etal11}M. Rotondo, J. A. Rueda, R. Ruffini, and S.-S. Xue, Phys.
Rev. C \textbf{83}, 045805 (2011).

\bibitem {Web+etal14}F. Weber, G. A. Contrera, M. G. Orsaria, W. Spinella, O.
Zubairi, Mod. Phys. Lett. A \textbf{29}, 1430022 (2014).

\bibitem {GelTan16}F. Gelis and N. Tanji, Prog. Part. Nucl. Phys. \textbf{87},
1 (2016).

\bibitem {allor08}D. Allor, T. D. Cohen, and D. A. McGady, Phys. Rev. D
\textbf{78}, 096009 (2008).

\bibitem {GavGitY12}S. P. Gavrilov, D. M. Gitman, and N. Yokomizo, Phys. Rev.
D \textbf{86}, 125022 (2012).

\bibitem {VafVish14}O. Vafek, A. Vishwanath, Annu. Rev. Condens. Matter Phys.
\textbf{5}, 83 (2014).

\bibitem {KaneLM15}G. Kan\'{e}, M. Lazzeri, and F. Mauri, J. Phys.: Condens.
Matter \textbf{27}, 164205 (2015).

\bibitem {Olad+etal17}I. V. Oladyshkin, S. B. Bodrov, Yu. A. Sergeev, A.
I.\ Korytin, M. D. Tokman, A. N. Stepanov, Phys. Rev. B \textbf{96}, 155401
(2017).

\bibitem {Akal+etal19}I. Akal, R. Egger, C. M{\"u}ller, S. Villalba-Ch\'{a}vez, Phys.
Rev. D \textbf{99}, 016025 (2019).

\bibitem{LaserRev} G. V. Dunne, Eur. Phys. J. D \textbf{55}, 327 (2009); A. Di Piazza, C. M{\"u}ller, K. Z. Hatsagortsyan, and C. H. Keitel, Rev. Mod. Phys. \textbf{84}, 1177 (2012); G. Mourou and T. Tajima, Eur. Phys. J. Special Topics \textbf{223}, 979 (2014); G. V. Dunne, Eur. Phys. J. Special Topics \textbf{223}, 1055 (2014); B. M. Hegelich, G. Mourou, and J. Rafelski, Eur. Phys. J. Special Topics \textbf{223}, 1093 (2014).

\bibitem {FGS}D. M. Gitman, J. Phys. A \textbf{10}, 2007 (1977); E. S.
Fradkin, D. M. Gitman, Fortschr. Phys. \textbf{29}, 381 (1981); E. S. Fradkin,
D. M. Gitman, and S. M. Shvartsman, \emph{Quantum Electrodynamics with
Unstable Vacuum} (Springer-Verlag, Berlin, 1991).

\bibitem {Nikis79}A. I. Nikishov, in \emph{Quantum Electrodynamics of
Phenomena in Intense Fields}, Proceedings of the P.N. Lebedev Physics
Institute Vol. 111 (Nauka, Moscow, 1979), p. 153.

\bibitem {General1}N. D. Birrell and P. C. W. Davies, \emph{Quantum Fields in
Curved Space} (Cambridge University Press, Cambridge, England, 1982); A. A.
Grib, S. G. Mamaev, and V. M. Mostepanenko, \emph{Vacuum Quantum Effects in
Strong Fields} (Friedmann Laboratory, St. Petersburg, 1994).

\bibitem {Ruf10}R. Ruffini, G. Vereshchagin, and S. Xue, Phys. Rep.
\textbf{487}, 1 (2010).

\bibitem {Dunn04}G. V. Dunne, in \emph{From Fields to Strings:
Circumnavigating Theoretical Physics}, edited by M. Shifman, A. Vainshtein,
and J. Wheater (World Scientific, Singapore, 2005).

\bibitem {DunnH98}G. Dunne and T. Hall, Phys. Rev. D \textbf{58}, 105022 (1998).

\bibitem {GusSh99}V. P. Gusynin and I. A. Shovkovy, Can. J. Phys. \textbf{74},
282 (1996); J. Math. Phys. (N.Y.) \textbf{40}, 5406 (1999).

\bibitem {GalN83}G. V. Galtsov and N. S. Nikitina, Zh. Eksp. Teor. Fiz.
\textbf{84}, 1217 (1983) [Sov. Phys. JETP \textbf{57}, 705 (1983)].

\bibitem {GiesK17}H. Gies and F. Karbstein, J. High Energy Phys. 03 (2017) 108.

\bibitem {GavGit17}S. P. Gavrilov and D. M. Gitman, Phys. Rev. D \textbf{95},
076013 (2017).

\bibitem {x-case}S. P. Gavrilov and D. M. Gitman, Phys. Rev. D \textbf{93},
045002 (2016).

\bibitem {L-field}S. P. Gavrilov and D. M. Gitman, Phys. Rev. D \textbf{93},
045033 (2016).

\bibitem {x-exp}S. P. Gavrilov, D. M. Gitman, and A. A. Shishmarev, Phys. Rev.
D \textbf{96}, 096020 (2017).

\bibitem {cr-regime1}H. Gies and G. Torgrimsson, Phys. Rev. Lett. \textbf{116},
090406 (2016).

\bibitem {cr-regime2}H. Gies and G. Torgrimsson, Phys. Rev. D \textbf{95}, 016001
(2017).

\bibitem {AdoGavGit17}T. C. Adorno, S. P. Gavrilov, and D. M. Gitman, Int. J.
Mod. Phys. A. \textbf{32}, 1750105 (2017).

\bibitem {instantons}I. K. Affleck, O. Alvarez, and N. S. Manton, Nucl. Phys.
B\textbf{197}, 509 (1982); G. V. Dunne and Ch. Schubert, Phys. Rev. D
\textbf{72}, 105004 (2005); G. V. Dunne, Q.-H. Wang, H.
Gies, and Ch. Schubert, Phys. Rev. D \textbf{73}, 065028 (2006); G. V. Dunne and Q.-H. Wang, Phys. Rev. D \textbf{74},
065015 (2006).

\bibitem {Karb17}F. Karbstein, Phys. Rev. D \textbf{95}, 076015 (2017).

\bibitem {BrGavGitIv} A.I. Breev, S.P. Gavrilov, D.M. Gitman, and D.A. Ivanov, arXiv:1903.06832.
\end{thebibliography}
\end{document}